# Insights into spatio-temporal dynamics during shock – droplet flame interaction

Gautham Vadlamudi [1], Akhil Aravind [1], Saini Jatin Rao [1], and Saptarshi Basu [1,2]*

*: Corresponding author email: sbasu@iisc.ac.in

**Affiliations:**

[1] Department of Mechanical Engineering, Indian Institute of Science, Bangalore 560012, India

[2] Interdisciplinary Centre for Energy Research (ICER), Indian Institute of Science, Bangalore 560012, India

## Abstract

The study comprehensively investigates the response of a combusting droplet during its interaction with a high-speed transient flow that is imposed by a coaxially propagating blast wave. The blast wave is generated using a specially designed unique miniature shock generation apparatus that generates blast waves using the wire-explosion technique which facilitates a wide range of shock Mach number ($1.03 < M_s < 1.8$). The experiments are performed in two configurations: Open field blast wave and focused blast wave. The charging voltage and the configuration determine the shock Mach number ($M_s$) and flow characteristics. The flame is found to exhibit two major response patterns: partial extinction followed by re-ignition and full extinction. Simultaneously, the droplet also interacts with the flow imposed by the blast wave exhibiting different modes of response ranging from pure deformation, Rayleigh-Taylor piercing bag breakup, and shear-induced stripping. The KH instability is exhibited along the windward side interface of the droplet during the interaction with the blast wave decay profile which gets aggravated when the induced flow interaction ensues. Increasing the Mach number ($M_s > 1.1$) makes the droplet flame more vulnerable to extinction. However, the flame exhibits stretching and shedding, followed by re-ignition at lower Mach numbers ($M_s < 1.06$). In all cases, the flame base lifts off in response to the imposed flow, and the advection of the flame base interacting with the flame tip results in flame extinction. The entire interaction occurs in two stages: 1) interaction with the blast wave and the decaying velocity profile associated with it, and 2) interaction with the induced flow behind the blast wave as a result of the entrainment (delayed response). The criteria for partial and complete extinction of flame have been postulated which is in good agreement with the experiments.

## 1. Introduction

The ongoing progress in combustion systems for advanced propulsion, along with the numerical modelling of two-phase combustion, has spurred extensive research into the fundamental dynamics of shock wave interactions with multiphase flows. One of the important applications of flame–shock interaction includes Scramjet combustors that operate with high-speed supersonic intake air as the oxidizer. These combustors contain complex flow features including shock waves, expansion fans, boundary layers, etc, and the interaction of the flame with these features becomes the key in the development of scramjet engines (Roy and Edwards 2000). Liquid fuel compared to gaseous fuels offers an intrinsic advantage of higher volumetric energy density which is desirable for propulsion applications (Patten et al. 2023, Anderson and Schetz 2005). Thus, the interaction between individual combusting fuel droplet (formed from spray atomization) and the shock structures becomes essential for the combustion process.

Another application where the shock-flame interactions are relevant is the detonation-based engines which offer higher theoretical efficiencies compared to deflagration engines. Studies have demonstrated that utilizing detonations, rather than deflagrations, to release the fuel's energy can potentially enhance the specific impulse of an air-breathing engine (Kashdan et al. 2004). The phenomena of the deflagration to detonation transition (DDT) and shockwave–induced combustion in liquid fuels are highly relevant in such applications (Patten et al. 2023, Dyson et al. 2022, Kashdan et al. 2004, Wei et al. 2017, Ciccarelli et al. 2010). Furthermore, fire-fighting applications also employ explosives for extinguishment of large-scale fires or oil well fires (high fuel flow rate), which are harder to extinguish by conventional methods (Yoshida et al. 2024, Chan et al. 2016). This blast extinguishing method blows off the fire due to the fluid dynamical effects of the blast wave and the flame shock interactions are essential in such applications as well.

The interaction of flame with shock wave/acoustic wave is associated with various flame instabilities (Maley et al. 2015, Jiang et al. 1997, Tyaktev et al. 2020, Khokhlov et al. 1999) which affect the configuration and propagation of flames especially in confined chambers. Wei et al. have also shown that the flame shock interaction plays a significant role in flame Heat release rate enhancement due to flame distortions (Wei et al. 2017). Thomas et al. (2001) showed enhancement in combustion driven by chemi-acoustic interactions and gas-dynamics effects when laminar flame bubbles were perturbed successively by incident and reflected shock waves. The shock-flame interactions show that the spherical bubble transforms into a toroidal shape due to vorticity generation (Picone and Boris 1988, Ju et al. 1988). It is also shown that as the shock strength is increased, the local gas temperature and pressure increase which results in a transition to detonation (DDT) in the immediate vicinity of the reaction front. Researchers like Wei et al. (2017), Ciccarelli et al. (2010) studied flame shock interaction in the confined chamber by obstructing the flame propagation using a perforated plate. Different modes of combustion were reported which involve flame acceleration, DDT, and autoignition. Dong et al. (2008) numerically investigated the interaction of spherical flame bubble with incident planar shock wave and showed that the hydrodynamic processes play a more important role than the chemical processes. Numerical investigations by Khokhlov et al. (1999) showed that interaction between shocks and flames is responsible for maintaining highly turbulent flame through Richtmyer Meshkov (RM) instability and is essential for DDT to occur by creating local hotspots. The results show that the RM instability contributes to the flame stability and enhancement of mixing.

A typical shock wave is a shock front that is followed by constant flow properties (steady flow) and a blast wave is a shock front with exponentially decaying flow properties (unsteady) behind it, which is characterized by a temporally decaying Mach number (Apazidis and Eliasson 2019). It has been established that the blast wave initially exhibits a pseudo-steady shock front behaviour and shows minimal variation in shock Mach number and thus, minor change in properties. However, as it propagates further, it exhibits non-linearity, thus, deviating from the linear steady shock behaviour and it eventually approaches a weak blast wave limit, thus behaving as an acoustic wave (Almustafa and Nehdi 2023). Wei and Hargather (2021) developed scaling based on dimensional analysis for estimating the shock trajectory. It is shown that for strong shock limit ($M_s > 5$) the temporal evolution of shock radius approximately follows a power law with an exponent 2/5 and for weak shock limit ($M_s \rightarrow 1$), the shock radius is a linear function of time. The decaying blast wave is shown to follow strong shock limit in early stages of high energy blast wave (only valid for nuclear scale explosions) and gradually transition into intermediate phase finally asymptotically approaching acoustic limit in the far field (Díaz and Rigby 2022). In the current experiments, the Mach number of blast wave is in the range of $(1.02 < M_s < 1.5)$ which is in the intermediate transition regime near the weak blast wave limit.

A blast wave is characterized by a discontinuity in properties such as peak overpressure which continuously decays away from the blast wave. As the blast wave propagates, the pressure at a given point temporally decays initially from the maximum value (positive phase), which further decays below zero thus, attaining negative pressures in later stages. Regardless of source geometry it is observed that, the blast wave shock front eventually attains notably spherical form after propagating sufficient distance (Almustafa and Nehdi 2023). Researchers like Taylor (1950), Sedov (1957) showed that a self-similar solution is only valid for the strong shock regime (i.e., $1/M_s \rightarrow 0$) that occurs during the initial stages of the blast wave. Later for intermediate shock strengths, a departure from the self-similar solution due to counterpressure effects is observed, which is accounted for in the perturbation solution using the approximate linear velocity profile solution by Sakurai et al. (1956) and quasi-similar solution of Oshima (1960).

In the analysis by Rae (1965) and Lee (1965) a power-law density profile behind the blast wave whose exponent can be obtained from mass integral. Bach and Lee (1970) showed that the particle velocity profile behind the blast wave can be obtained using density profile and the mass conservation differential equation. Thus, the obtained density and velocity profiles can be used to obtain pressure profile using the momentum equation. Using the energy integral, the shock decay coefficient dependence with respect to the Mach number ($M_s$) can be obtained in the form of first order differential equation from which shock profile can be estimated. This approximate analytical solution based on power-law density profile assumption provided by Bach and Lee (1970) is observed to be in good agreement with the exact numerical solution by Goldstine and von Neuman (1963) [Goldstine, H. and von Neumann, J., "Blast Wave Calculation," Collected Works of J. von Neumann, Vol. VI, Pergamon Press, Oxford, England, 1963, pp. 386-412; also Macmillan, New York] even at the low shock strength regime ($M_s \rightarrow 1$) along with strong shock limit ($1/M_s \rightarrow 0$).

Current experiments focus on the flame and droplet response during their interaction with the flow imposed by the blast wave. Droplet combustion has been a significant area of research due to its relevance in spray combustion across a multitude of applications such as power generation, basic transportation to aero propulsion. Although droplet combustion cannot be directly extended to practical results for reacting sprays, it provides insights into the local phenomena such as flame stabilization, forward extinction, blow-off, pollutant formation, etc under well-controlled conditions (Williams 1973). For a quiescent burning droplet, fuel vaporizes at the droplet surface and a diffusion flame is formed at the stoichiometric plane enveloping the droplet. The $d^2$-law law has been established as the droplet regression model and researchers have investigated droplet and researchers have investigated droplet combustion in both micro-gravity and gravitational environments (Law 1982, Hara and Kumagai 1994, Huang and Chen 1994). It has been reported that the relative motion between the droplet and surrounding gases alter the vaporization characteristics and flame configuration. The flame undergoes local extinction at the forward stagnation point and transitions into the droplet wake under the influence of external flow and such investigations provide insights into local extinction events which occur in spray combustors. The combustion experiments have been conducted by Balakrishnan et al. (2001) to investigate the enveloped flame shape characteristics in a mixed convective environment. Researchers have also conducted pendant droplet experiments under different externally imposed flow and studied the aspects like droplet regression rate, internal boiling, secondary atomization which enhances the heat release rate of the flame (Basu et al. 2014, Guerieri et al. 2015, Guerieri et al. 2017). Researchers like Pandey et al. (2020), Vadlamudi et al. (2021, 2023), and investigated the effect of continuously varying external flow on droplet combustion in free fall configuration using drop tower experiments. The flame stabilization criteria, flame evolution, flame topology have been investigated for the Reynolds number range of $0 < Re < 200$. Pandey et al. (2021) and Thirumalaikumaran et al. (2022) performed pendant droplet experiments to investigate the flame shedding phenomena based on circulation buildup mechanism.

In current experiments, flame – shock interactions are investigated by studying the interaction of a combusting droplet with a temporally decaying blast wave (that propagates coaxially). Oshima (1960) investigated the blast wave generated through wire-explosion technique, in which a fine metal wire is burned out explosively by discharging large voltage pulse through it to generate a cylindrical blast wave. Sharma et al. (2021, 2023a, 2023b) and Chandra et al. (2023) conducted shock – droplet interaction experiments using exploding wire technique using a miniature shock tube. It was shown that during the shock interaction, the droplet dynamics broadly occur in two stages: initial deformation and subsequent breakup. However, because of experimental limitations, normal shock assumptions were considered for the flow behind the blast wave and the flow has not been fully characterized. However, the droplet dynamics and breakup criteria based on a both Weber number and Ohnesorge number and showed that shear-stripping mode of breakup occurs for droplet diameters significantly higher than the order of magnitude of the wavelength of the KH waves. The droplet breakup dynamics were shown to be qualitatively similar even for liquid metal droplets if the surface oxidation can be restricted (which leads to flake-like breakup).

Researchers like Chan et al. (2016) studied the interaction of shock tube exhaust flow with non-premixed propane flame (perpendicular to the shock tube axis, transverse direction) and showed different types of interaction modes. The shock is generated using a conventional compressed air-driven shock tube generating low Mach numbers ($M_s < 1.5$) and is allowed to interact perpendicularly with a non-premixed jet flame. The non-premixed propane flame exhibited extinction either during interaction with the shock wave or with the blast wind vortex, with reignition occurring only when flame is in the far field location. Current experiments investigate the interaction of shock wave with a combusting droplet (in longitudinal direction) where both the flame dynamics as well as the droplet response has been studied. Since, the experiments are conducted in the longitudinal direction, a wide range of phenomena such as the flame shedding, base lift-off, reattachment, partial extinction, reignition and full blowout are observed.

In droplet combustion, the droplet flame sustains on the available fuel vapor which is dependent on the vaporization rate at the droplet. Thus, the shock interaction with the droplet flame becomes a unique process where the shock interaction can affect the droplet dynamics as well as the fuel availability which alters the flame dynamics. Unlike other shock – flame interaction studies in the literature, this results in simultaneous response of both droplet and flame independently to the imposed shock as well as influencing each other. In this study, a unique miniature shock generation apparatus is used to generate blast waves to achieve a wide range of Mach numbers ($1.02 < M_s < 1.6$) to study the interaction of blast waves with a combusting droplet. Present work focuses on the fluid dynamic (flow) aspect of the flame response during the interaction with the flow imposed by the blast wave. Building on to the previous works, attempt has been made to characterize the flow features, and velocity scales imposed by the blast wave that is generated using the wire explosion technique. The decaying profile behind the blast wave has been considered in current study instead of the normal shock assumptions and different sets of experiments were performed to gain better understanding of the flow features present behind the blast wave. Additionally, the time scales involved for the two stages of the droplet dynamics were also explored in current work which have not been discussed in the literature. Furthermore, the interaction of combusting droplet with flow imposed by blast wave will be a unique and insightful addition to the existing shock – flame interaction literature as well as the droplet combustion literature.

## 2. Experimental Methodology

A specially designed shock tube apparatus is used in current experiments where the shockwave is generated using the exploding wire technique. This technique allows the Mach number ($M_s$) to be controlled by altering the voltage applied for the wire explosion. The range of Mach numbers ($M_s$) achieved through this setup in current experiments is from 1.03 to 1.6. Researchers like Liverts et al. (2015) and Sembian et al. (2016) give a detailed overview of the exploding wire technique and its applications in shockwave generation. Compared to the diaphragm-based shock tubes, this technique allows a smaller test facility size, better ease of operation, and the generation of a wide range of $M_s$ (Sembian et al. 2016).

Figure 1 shows the schematic of the experimental setup consisting of a shock generation setup and flow visualization camera to study the interaction of the blast wave with the combusting fuel droplet (in the longitudinal direction). The figure shows the shock generation consisting of an electrode chamber that is enclosed with a cover plate. A 2kJ pulse power system (Zeonics Systech, India Z/46/12) that can discharge a 5μF capacitor is used to provide a high-voltage pulse across the electrodes. To achieve the wire explosion, a copper wire of 35 SWG is placed over the cover plate in electrical contact with the two electrodes of the electrode chamber, which is, in turn, connected to a high-voltage power supply. An external BNC 745 T digital delay generator is used to synchronize and trigger all the recording devices and the shock generator by sending a trigger in the form of a TTL signal at pre-specified time delays. During the experiment, the capacitor is initially charged to a desired energy level required to generate a specific shock Mach number, and the charging circuit is cut off. A 1kV trigger signal is provided to the variable spark-gap switch as soon as an external TTL trigger signal is received to close the discharging circuit (containing the electrodes and the copper wire). As soon as the trigger signal is received and the discharging circuit is closed, the high-voltage pulse discharges through the electrodes and the copper wire, resulting in the rapid Joule's heating and vaporization of the thin copper wire. This generates a cylindrical blast wave, and this technique of producing a shockwave is known as the exploding-wire technique.

A n-dodecane droplet of ~ 2 mm diameter (d) is suspended in pendant mode at a distance of 365 mm from the copper wire using a quartz rod of 0.4 mm diameter. A heater coil attached to a linear solenoid is used for igniting the droplet. The experiments were performed in two different configurations: Open-field (figure 1a) and Shock tube focusing (figure 1b). In the open-field configuration, the wire explosion occurs in the open, and a cylindrical blastwave propagates radially outward from the exploding wire (see Figure 1a). In the latter case, a rectangular flow channel (shock tube) is firmly mounted onto the cover plate of the electrode chamber (using bolts) to direct and focus the blast wave along the shock tube (see figure 1b). Two different rectangular flow channels (made of a high impact strength material - polycarbonate) designed with the internal cross-sectional dimensions of 100 mm × 20 mm and 40 mm × 20 mm (each of length 330 mm) were used as shock tubes in the experiments to obtain different $M_s$ (figure 1c). The shock tube is firmly fixed to the electrode chamber cover plate (using nut-bolt arrangement) to minimize leakage of the compressed air. It is to be noted that in both Open-field and shock tube focusing configurations, the droplet location from the wire is maintained at the same distance.

In the Open-field configuration (no shock tube), the cylindrical blast wave propagates radially outward away from the origin (copper wire). Whereas in the presence of a shock tube, the cylindrical blast wave transforms into a planar shock inside the rectangular column due to the geometry of the flow channel (Sembian et al. 2016). Due to the experimental limitations of igniting the fuel droplet, the pendant fuel droplet had to be placed at a distance of 35 mm from the shock tube exit (to allow for the movement of the heater). Thus, when the planar propagating blast wave from the wire explosion along the rectangular channel exits the shock tube, it expands into the open ambient similar to an expanding cylindrical blast wave as it reaches the combusting droplet.

Different charging voltages for the capacitor from 4 kV to 10 kV were used in the current experiments for each of the three experimental configurations: open-field and configuration with shock tube of two different dimensions (see figure 1c). The synchronization of the droplet ignition, blast wave interaction, and simultaneous high-speed recording of flow visualization is achieved using BNC 745 T digital delay generator within nanosecond accuracy. The droplet ignition is achieved using an Arduino circuit which releases the heater coil mounted on a pull-type linear solenoid to ignite the droplet. After a specified time delay, the linear solenoid is actuated, and thus, the heater coil is retracted away from the droplet, simultaneously sending an output trigger TTL signal. As soon as the BNC digital delay generator receives this input TTL signal, it triggers all the devices connected to it after the predetermined time delay intervals. Since the phenomena observed during the experiments, i.e., blast wave propagation, flame interaction, and droplet break up, occur at microsecond time scales, the integration and synchronization of different components are crucial for experimentation. Three different sets of experiments were performed using high-speed Schlieren imaging, high-speed Mie-scattering imaging for flow visualization, high-speed OH*Chemiluminescence, and high-speed shadowgraphy imaging (side view) for the droplet breakup.

A Schlieren system is used for the flow visualization using two spherical concave mirrors (1.5 m focal length) and a high-speed non-coherent pulse diode laser of 640 nm wavelength (Cavitar Cavilux smart UHS, 400 W power) along with a knife-edge. The light beam emitted by the Cavilux laser source is transformed into a point light source using a variable round iris aperture (Holmarc SSID-25) which is placed at the focal length of the first concave mirror (M1). This forms a parallel light beam which is directed towards the test section, for recording the shock-droplet flame interaction. Another spherical concave mirror (M2) is placed on the other side, which is used to focus the incoming parallel light beam from the test section into a point at its focal length. A knife edge is used to block the incoming converging light from M2 (placed at the focal length of M2 mirror) to visualize the density gradient variation in the flow field during the shock interaction phenomena. This Schlieren arrangement facilitates the visualization of complex wave structures due to the blast wave and the flow features around the droplet flame in terms of density gradients (see figure 1a,b). A high-speed Photron SA5 camera is used for recording the Schlieren imaging at the acquisition rate of 75000 fps, 256 × 312 pixels frame (at a pixel resolution of 0.2375mm/px) at which the pulse laser is synchronized. Schlieren imaging is used to obtain the shock Mach number and other flow features.

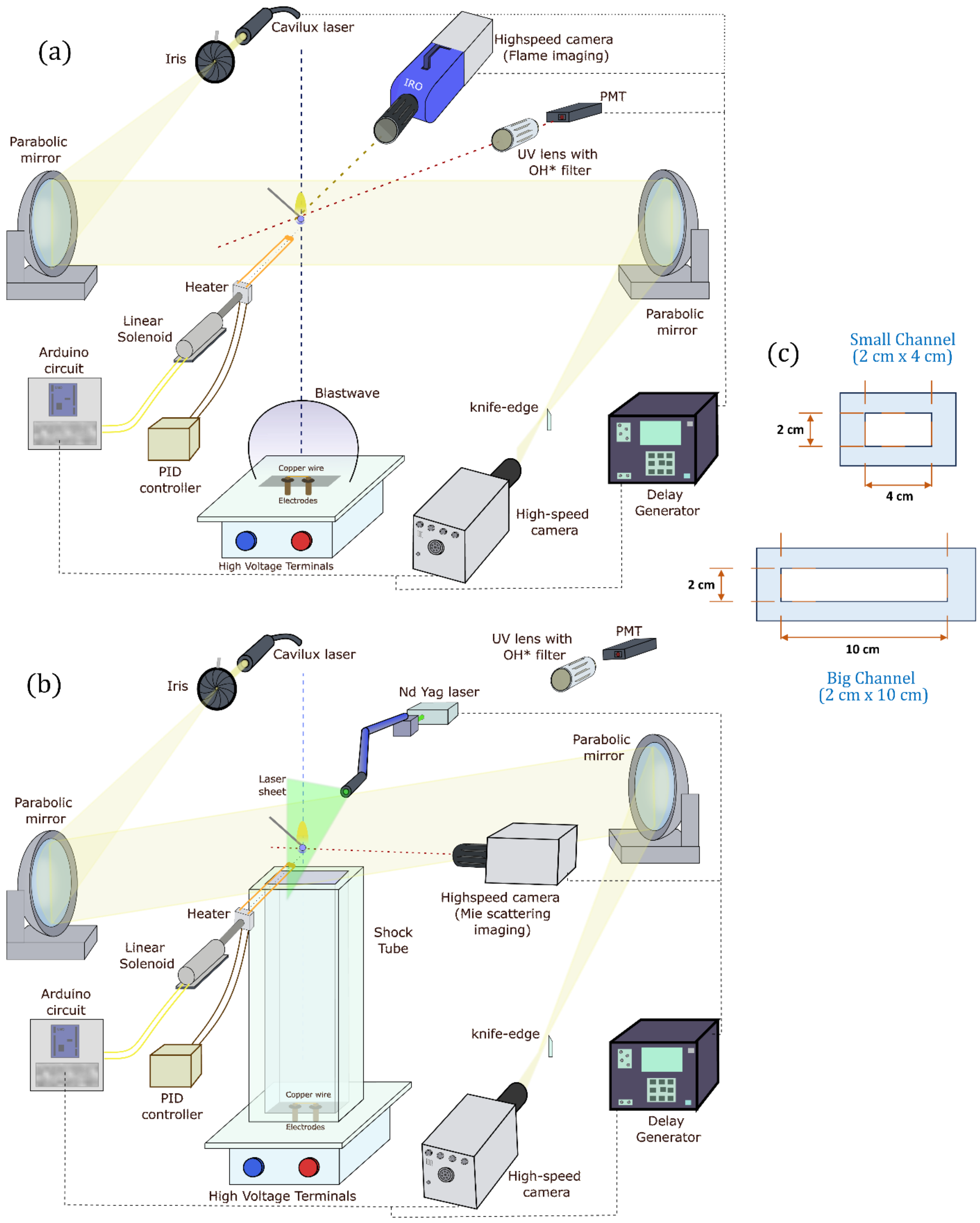

**Figure 1.** Experimental setup**:** (a) Simultaneous high-speed Schlieren and flame imaging of the shock – droplet flame interaction performed for both Open-field blast wave and shock tube focusing configurations (only open-field configuration is shown in the figure). (b) Simultaneous high-speed parallax Schlieren and Mie-scattering flow visualization of the central plane in shock tube focusing configuration. (c) The cross-sectional dimensions of the two shock tube channels.

It is to be noted that the quartz rod is placed at a skewed angle to the focal plane of the Schlieren camera, so that only the tip of the quartz rod (the droplet portion) is in the focal plane of the Schlieren camera. The quartz rod is chosen over metal wire for holding the droplet in pendant mode to minimize the heat conduction effects as well as heterogeneous nucleation effects inside the droplet (which leads to internal boiling and bubble formation inside the droplet). The quartz rod diameter is also chosen to be as small as possible ($d_q$~0.4 mm) to minimize the effects of the quartz rod.

For high-speed shadowgraphy, a high-speed Cavilux pulse laser is used as the backlight by collimating the laser beam into a parallel light beam. A high-speed Photron SA5 camera synchronized with it is aligned in-line with it for imaging. The droplet shape dynamics are recorded using high-speed shadowgraphy (at 75000 fps, 256 × 312 pixels frame at a pixel resolution of 0.05mm/px) to study its deformation and breakup phenomena.

High-speed flame imaging (OH*Chemiluminescence) is also performed using a high-speed star Lavision SA5 Photron camera, coupling it with a high-speed intensifier (HS IRO, Lavision; IV Generation) alongside a Nikon Rayfact PF10445MF-UV lens and an OH*Band pass filter (~ 310 nm). The focus of the current flame imaging experiments is limited to studying the flame response to the external flow imposed by the blast wave. The flame imaging is performed at 20000 fps, 256 × 312 pixels frame (at a pixel resolution of 0.081mm/px).

The Mie-scattering imaging is primarily used for visualizing the induced flow (vortical structures) behind the blast wave (see figure 1b). The Mie scattering flow visualization is performed by using a high-speed dual-pulsed Nd;YLF laser with a wavelength of 532 nm and pulse energy of 30 mJ per pulse as an illumination source. The cylindrical output beam of the laser (5 mm diameter) is converted into a thin sheet of 1 mm thickness using sheet-making optics. The diethyl hexa sebacate (DEHS) (1-3 μm, density $\rho$ = 912 $kg/m^3$) oil droplets are used as the seeding particles. Initially, the seeder particles are filled inside the shock tube by closing the open end with a lid, and just before triggering the system, the lid is removed from the shock tube exit, and then the system is triggered. A Lavision SA5 camera, which is synchronized with the 532 nm laser via a programmable tuning unit (PTU), operates the system in a single-frame mode at an acquisition rate of 6000 fps and imaging pixel resolution of 1024 × 1024 pixels. The optical axis of the camera is aligned orthogonally to the plane of the laser sheet. A 532 nm band-pass filter is coupled with the camera lens for flow visualization to capture the Mie-scattered light from seeder particles in the test section. The double-frame Mie-scattering imaging is also performed at 3000 fps with a double-pulse having 10 μs separation. The PIV Lab module in Matlab has been used for the post-processing to obtain the order of magnitude of the velocity scale of the induced flow vortical structures behind the blast wave. Due to the experimental limitations, only the order of magnitude of the velocity of vortical structures has been obtained, and they have not been used to reconstruct the velocity of the field. The Mie-scattering experiments were only performed for both the focused-shock configurations (with the shock tube).

Along with the Schlieren imaging, high-speed imaging was also performed simultaneously at 6250 fps with a pixel resolution of 320 × 512 pixels using a Phantom MIRO camera. However, this flame imaging is for qualitative purposes to get a better insight into the flame dynamics and droplet breakup during the interaction with the blast wave. Since the interaction is associated with a very short time scale compared to the 6250fps acquisition rate, it results in motion blurring in the flame images (see figure 1a). The Schlieren imaging is also performed simultaneously along with the Mie-scattering flow visualization and high-speed

OH* chemiluminescence imaging (see figure 1b). Shadowgraphy imaging is performed simultaneously with flame imaging.

All the experimental imaging (Schlieren, shadowgraphy, flame imaging) were performed for all the three configurations: i.e., Open-field, 2 cm × 10 cm c/s channel and 2 cm × 4 cm c/s channel. Three experimental runs were conducted for each case in all the experimental setups. During the experiments, the charging voltage was varied from 5 kV to 10 kV and between 4 kV to 8 kV for Open-field and with-shock tube configurations respectively to control the shock strength. The Shock Mach number ($M_s$) is observed to vary with both the change in shock tube configuration as well as the charging voltages. The Mach number ($M_s = V_s / c$) is obtained using the blast wave propagation velocity ($V_s$) which is obtained by measuring the distance travelled by the cylindrical blast wave between two consecutive frames recorded at 75000 fps along the centreline. The blast wave velocity is observed to decrease as it propagates downstream, which shows diminishing shock strength as the blast wave expands into the ambient after exiting the shock tube.

The bright flame tip, as well as the low-density hot gases at the flame, are visible in the Schlieren imaging. Thus, the flame base and flame tip can be visually tracked using high-speed Schlieren imaging. Additionally, Schlieren experiments (75000 fps) were also performed with a light source illuminating the test section only in three out of four subsequent frames (by using a logic gate circuit, shown in supplementary Figure S4). This allowed for better visualization of the flame dimensions (especially the bright tip) alongside the Schlieren imaging that shows the density gradient contrast due to the presence of hot gases at the flame. This data is corroborated using the simultaneous OH* Chemiluminescence flame imaging (at 20000 fps) to obtain the flame base and flame tip locations during its interaction with the shock. The flame images have been thresholded using the Otsu thresholding technique (in-built ImageJ). Otsu's thresholding algorithm computes a solitary intensity threshold (If) to partition all pixels within an image into two categories: foreground and background. This threshold (If) is established by minimizing the intra-class intensity variance or, alternatively, by maximizing inter-class variance. Thus, the flame boundary can isolated to obtain the flame dimensions and flame base lift-off. Time series data of Simultaneous OH* Chemiluminescence has been gathered using a Hamamatsu photomultiplier tube (H11526-110-NF) at a sampling rate of 75,000 Hz. Nonetheless, this data is beyond the scope of the present investigation and has not been utilized in the data analysis.

## 3. Results and Discussions

### 3.1. Flow characterization

After the system is triggered, the blast wave generated travels away from the electrodes towards the combusting pendant droplet coaxially. The blast wave travels from its origin and reaches the droplet flame location at $t = t_s$. The blast wave is visualized through Schlieren imaging as it passes by the droplet flame. The flame dynamics are observed to drastically vary with the shock tube configuration (Open-field vs focused) and charging voltages. The shock Mach number ($M_s$) is observed to decay temporally as the blast wave propagates against the quiescent ambient medium. However, all of the cases are characterized by the value of the Mach number ($M_s$) near the location of the droplet that is measured along the centreline. $M_s$ obtained at the droplet is observed to increase with the increase in the initial charging voltage as well as the extent of the focusing. Higher $M_s$ were obtained for the same charging voltage as the confinement of the blast wave was increased (using a shock tube). The Open-field blast wave showed the lowest $M_s$ values as the blast wave expanded outwards in all directions from the point of the wire blast. However, in the case of blast focusing (with shock tube), the blast wave is planar inside the shock tube flow channel (Sembian et al. 2016) and starts to expand only after exiting the shock tube. After the blast wave exits the shock tube and propagates downstream, an induced flow is observed behind the blast wave, which exits the shock tube and interacts with the combusting droplet after a time delay.

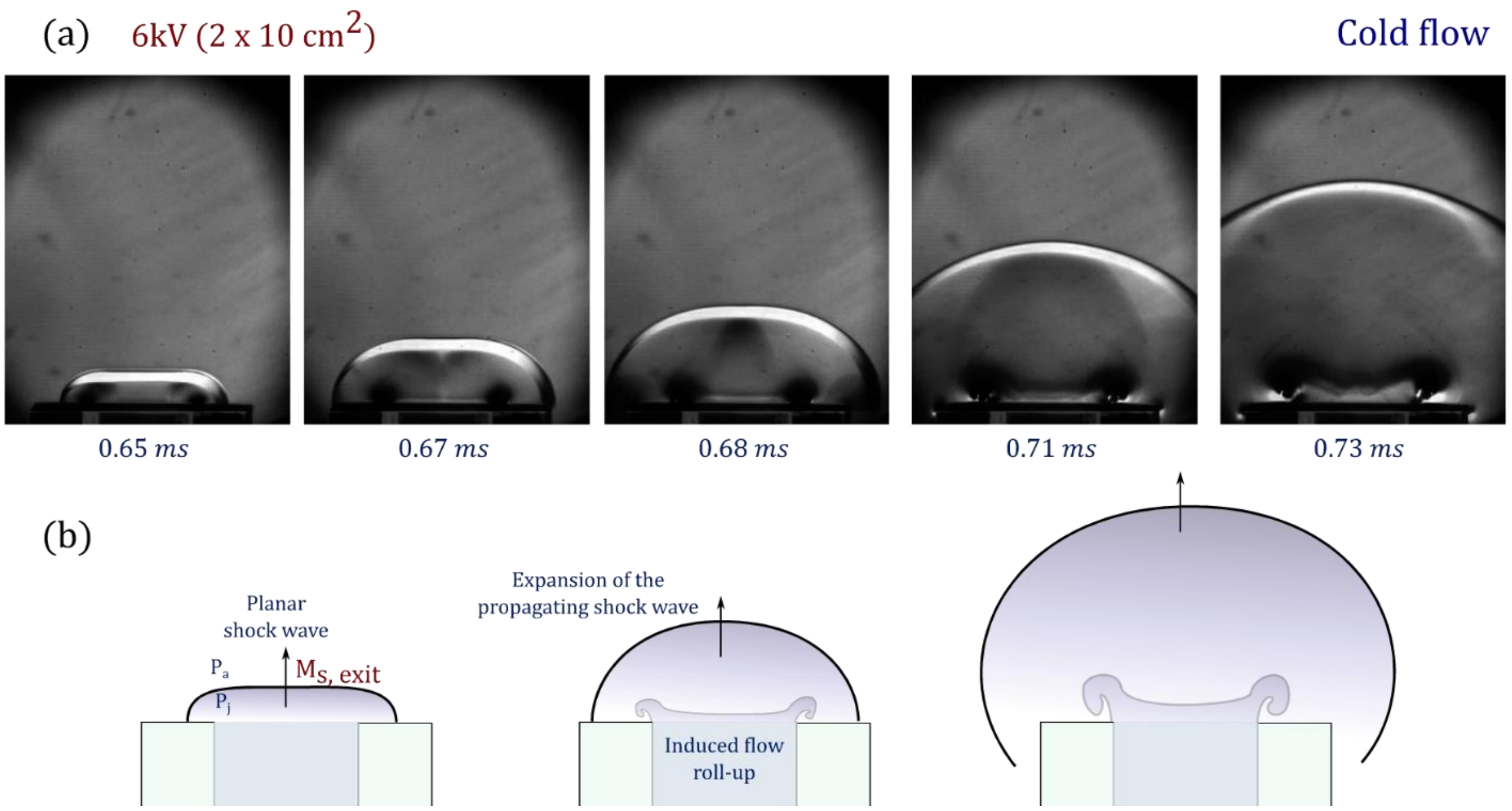

**Figure 2.** (a) Time series of the Schlieren imaging of the blast wave propagation near the shock tube exit. (b) Schematic of the blast wave propagation near the shock tube exit.

Figure 2a shows the time series of the flow exiting the shock tube channel with an expanding blast wave. Figure 2b gives a schematic of the flow. As the planar propagating blast wave inside the shock tube exits into the quiescent ambient, the blast wave expands outward similar to a radially expanding blast wave exhibiting significant curvature as it propagates. This radial expansion results in a continuous temporal decay of the shock strength and $M_s$ as the blast wave propagates further. Since the blast wave is almost planar at that instant as it exits the shock tube, the shock Mach number at the exit ($M_{s,e}$) is measured and normal shock relations can be used to evaluate the characteristics of the jet upstream of the blast wave. Table 1 gives the details of the initial upstream flow that is exiting the shock tube. On the other hand, in the case of Open-field configuration, a cylindrical blast wave propagates radially outward away from the copper wire having Mach numbers ($M_s$) of 1.03, 1.05, 1.055, 1.065 for the charging voltages of 5kV, 7kV, 8kV and 10kV respectively. Since, the droplet location from copper wire is far greater than the copper wire dimensions (w) i.e., $L_t + L_d >> w$, the copper wire can be assumed as a point source when the blast wave reaches the droplet (spherical blastwave assumption). Thus, the flow imposed by the blast wave in either of the configurations has to be characterized to understand the interaction with the droplet flame.

Using a similar methodology as Bach and Lee (1970) (which is valid for the $M_s$ range of current experiments i.e., $1.02 < M_s < 1.6$), a power-law density profile can be assumed behind the blast wave, and the velocity profile can be obtained from the mass conservation differential equation.

Initially, defining the non-dimensional parameters for velocity and density as follows:

$$\phi(\xi,\eta) = \frac{u(r,t)}{\dot{R}_s(t)} \quad (3.1)$$

$$\psi(\xi,\eta) = \frac{\rho(r,t)}{\rho_o} \quad (3.2)$$

where, $\theta(\eta) = \frac{R_s \ddot{R}_s}{\dot{R}_s}, \xi = \frac{r}{R_s(t)}, \eta = \frac{{c_o}^2}{{\dot{R}_s}^2} = \frac{1}{{M_s}^2}, y(\eta) = (R_s/R_o)^{j+1}$

The conservation of mass for unsteady 1D adiabatic motion of perfect gas behind the expanding blast wave is considered:

$$(\phi - \xi)\frac{\partial \psi}{\partial \xi} + \psi\frac{\partial \phi}{\partial \xi} + j\phi\frac{\psi}{\xi} = 2\theta\eta\partial\psi/\partial\eta \tag{3.3}$$

| | | Near the Shock tube exit | | | | Near the Droplet |
|---|---|---|---|---|---|---|
| | | Shock Mach number $(M_{s,e})$ | $\frac{P_j}{P_a}$ | Calculated Jet velocity $(U_j)$ | Jet Reynolds number $(Re_j)$ | Shock Mach number at the droplet $(M_s)$ |
| Big Channel (2 cm x 10 cm) | 4kV_Big | 1.09 | 1.22 | 50.30 | 6.45E+04 | **1.07** |
| | 6kV_Big | 1.28 | 1.77 | 147.24 | 1.89E+05 | **1.22** |
| | 7kV_Big | 1.43 | 2.21 | 210.05 | 2.69E+05 | **1.28** |
| | 8kV_Big | 1.49 | 2.41 | 235.05 | 3.01E+05 | **1.41** |
| | | | | | | |
| Small Channel (2 cm x 4 cm) | 4kV_Small | 1.16 | 1.43 | 90.22 | 1.16E+05 | **1.12** |
| | 6kV_Small | 1.38 | 2.07 | 191.23 | 2.45E+05 | **1.33** |
| | 7kV_Small | 1.56 | 2.68 | 265.89 | 3.41E+05 | **1.42** |
| | 8kV_Small | 1.6 | 2.83 | 282.68 | 3.62E+05 | **1.52** |

Table 1. Pressure ratio, jet velocity and Reynolds number upstream of the blast wave corresponding to different focused cases having different $M_s$ at the droplet.

$R_o$ is the characteristic explosion length, $R_s$ is the instantaneous shock radius, u is the velocity field, ρ is the density field and $c_o$ is the speed of sound. The boundary conditions at the shock front $\xi = 1$ are obtained from the standard normal shock relationship given by

$$\phi(1, \eta) = \frac{2(1-\eta)}{\gamma+1} \tag{3.4}$$

$$\psi(1, \eta) = \frac{\gamma+1}{\gamma-1+2\eta} \tag{3.5}$$

Thus, following Bach and Lee's (1970) formulation, the density profile can be assumed as follows:

$$\psi(\xi, \eta) = \psi(1, \eta)\xi^{q(\eta)} \tag{3.6}$$

The value of the exponent 'q' can be obtained using the mass integral obtained from the conservation of the mass enclosed by the blast wave at a given instant:

$$\int_0^1 \psi\xi^j d\xi = \frac{1}{j+1} \tag{3.7}$$

where the values j=0, 1, 2 are for planar, cylindrical, and spherical blast waves, respectively. Thus, the exponent of the density power-law is obtained to be:

$$q(\eta) = (j+1)[\psi(1, \eta) - 1] \tag{3.8}$$

The density profile can be obtained from Equations 3.5, 3.6, and 3.8 at any given instant for a specific Mach number ($M_s$). Thus, the mass conservation equation (3.3) can be rewritten using the density profile as follows:

$$\frac{\partial \phi}{\partial \xi} + (q+j)\left(\frac{\phi}{\xi}\right) = q + \frac{2\theta\eta}{\psi(1,\eta)}[1 + (j+1)\psi(1,\eta)\, ln\xi)]\frac{\partial \psi(1,\eta)}{\partial \eta} \tag{3.9}$$

Substituting the boundary condition at the origin as $\phi(0,\eta) = 0$, the velocity profile is obtained as:

$$\phi = \phi(1,\eta)\xi(1 - \Theta\, ln\xi) \tag{3.10}$$

where, $\Theta = \frac{-2\theta\eta}{\phi(1,\eta)\psi(1,\eta)} \frac{\partial \psi(1,\eta)}{\partial \eta}$

$\theta$ vs $\eta$ relation can be obtained in a similar methodology as shown by Bach and Lee (1970), and then Equation 3.10 can be used to obtain the temporally decaying velocity profile at a given location 'r' behind the blast wave for a given $M_s$.

The shock trajectory can be experimentally obtained by tracking the shock location along the centreline using high-speed Schlieren imaging. In the case of open-field blast wave, the shock is observed to propagate away from the wire location radially outward. Hence, the wire location can be considered as the origin (r = 0) for the open-field blast wave which can be considered as a spherical blast wave (j=2) at the droplet location with respect to copper wire dimensions ($L_t$ +$L_d$ >> w), see figure 3a. However, when the shock tube focusing is used, the blast wave is focused inside the shock tube along its length, which gets modulated as a planar blast wave as it propagates inside the shock tube (see figure 3a). However, when the blast wave exits the tube, it expands radially outward at the tube opening into the ambient atmosphere, similar to a blast wave. Experimentally, it is evident that the shape of the blast wave transitions from planar to cylindrical as it exits the rectangular shock tube opening. This is evident from the temporal variation of the radius of curvature ($R_s$) of the blast wave, which is observed to approach the instantaneous centreline distance (x) from the shock tube exit as the blast wave expands outward (see figure 3l). This implies that the blast wave tends to expand outward cylindrically, while approximately maintaining the centre of curvature at the shock tube exit location. That means the planar blast wave inside the shock tube transitions into a cylindrical blast wave with the shock tube exit location as its centre of curvature (origin). Hence, for simplicity, this expanding cylindrical blast wave can be assumed to be similar to a blast wave that has originated from the shock tube exit location as its virtual origin (r = 0) and is expanding radially outward into the ambient atmosphere, as shown in figure 3a. Thus, this expanding blast wave is assumed to be equivalent to the hypothetical cylindrical blast wave (j=1) originating from shock tube opening, and the instant when the blast wave exits the shock tube opening is considered to be the reference time ($t_v = 0$, virtual time), as shown in figure 3a. As shown in figure 3a, the length of the shock tube is $L_t$ and the droplet is placed at a distance of $L_d$ from the shock tube opening. Since the droplet position is maintained to be the same for both open and focused cases, the location of the droplet in the open field case is r = $L_t$ + $L_d$ (with r = 0 at the copper wire location).

Figure 3b,d shows the variation of Mach number along the shock trajectory location (mm) measured experimentally as the shock propagates for focused (i.e., big channel (B) and small channel (S)) and Open-field (Open) cases, respectively, designated based on different charging voltages (kV) and shock tube configuration. Higher charging voltages (kV) and smaller shock tube dimensions result in higher Mach numbers ($M_s$), and the open-field cases exhibited the lowest Mach numbers ($M_s$ < 1.08) for the same kilovolts (kV). The droplet location $L_d$ (in case of focused) and $L_t$ + $L_d$ (in case of open-field blast wave) are indicated in figure 3b,d with the help of a vertical orange dotted line. This implies the droplet and flame starts to experience the imposed flow of the blast wave only to the right side of this line. Figure 3c,e show the temporal variation of the Mach number ($M_s$) (measured experimentally) plotted against virtual time ($t_v$, ms). The theoretical blast wave trajectory can be obtained using the following expression (Bach and Lee 1970):

$$\frac{c_o t}{R_o} = -\frac{1}{2}\int_0^{\eta} \frac{y^{\frac{1}{j+1}} d\eta}{\theta \eta^{1/2}} \tag{3.11}$$

Where, $\theta(\eta)$ is shock decay coefficient and $y(\eta) = (R_s/R_o)^{j+1}$ is dimensionless instantaneous shock radius.

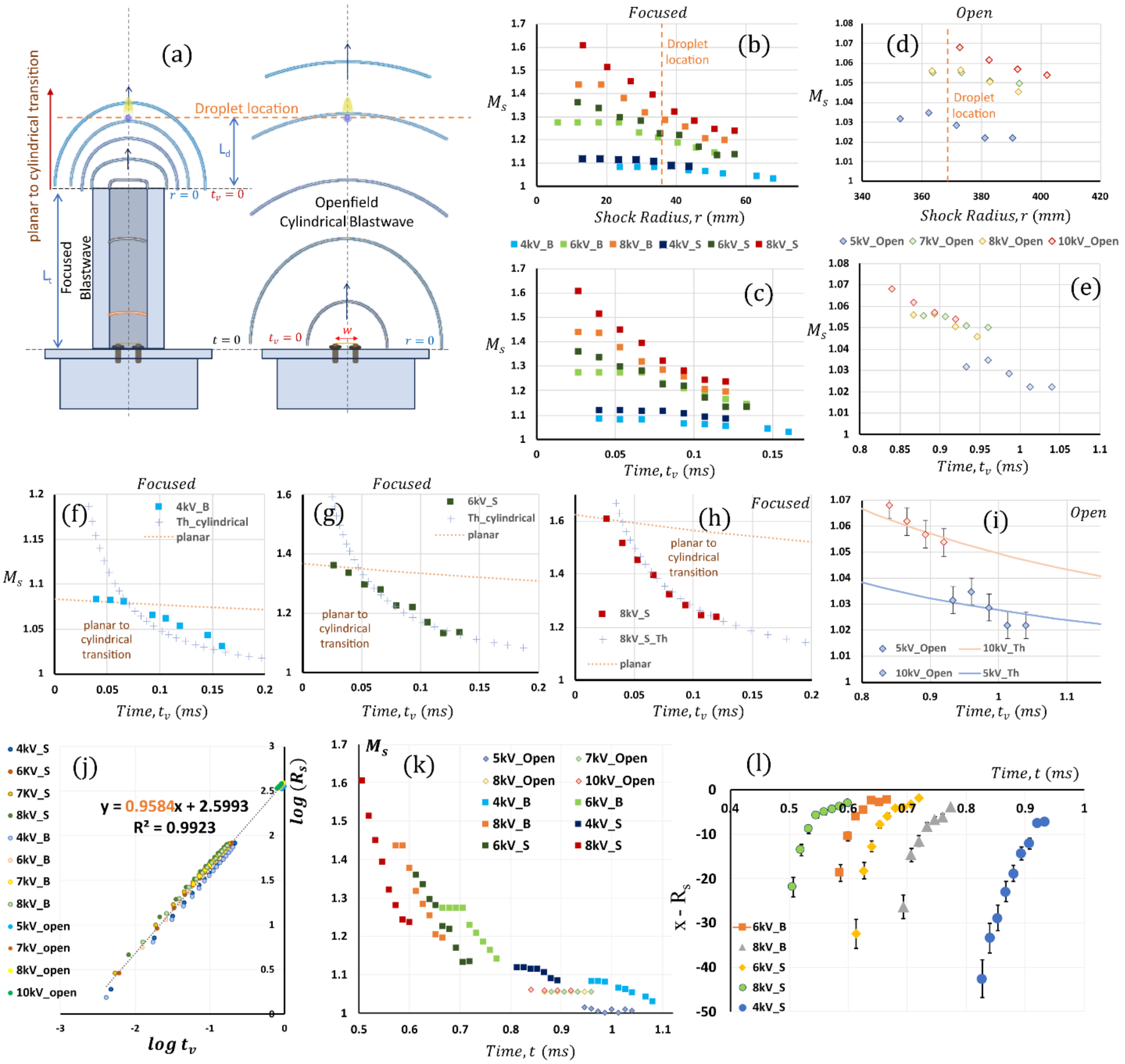

**Figure 3.** (a) Schematic of the blast wave propagation in open-field and focused configurations depicting the dimensions, droplet location and the cylindrical blast wave virtual origin (r=0), reference time ($t_v = 0$) for both the cases. (b,d) Variation of the Mach number along its trajectory (radial distance from origin, r=0) for focused and open-field blast wave cases respectively. 'kV' refers to the charging voltage for blast wave generation and 'B', 'S', 'Open' denote 'Big channel (2 cm × 10 cm c/s)', 'small channel (2 cm × 4 cm c/s)' and 'Open-field' configurations respectively. (c,e) Temporal variation of the Mach number along its for focused and open-field blast wave cases respectively. (f-i) Theoretical and experimental plots of Mach number ($M_s$) vs virtual time ($t_v$) of the blast waves generated for 4kV Big channel, 6kV, 8kV small channel and 5kV, 10kV Open-field configuration, respectively. (j) log-log plot of shock radius vs time; (k) Temporal

variation of the Mach number ($M_s$) for different cases plotted against time, t (in ms) with t = 0 at the time of the explosion; (l) Temporal variation of the deviation of instantaneous axial distance from shock tube exit, x and instantaneous radius of curvature ($R_s$) of the blast wave that is exiting the shock tube.

The parameters $\theta(\eta)$ and $y(\eta)$ were obtained by numerically solving the pair of first-order differential equations that are given by Bach and Lee (1970), which are obtained by substituting the profiles of different properties into the energy integral. The value of the characteristic explosion length ($R_o$) is dependent on the initial energy input to the blast wave explosion. However, in current experiments, the assumption of mass and energy conservation of Bach and Lee (1970) is not fully valid due to the entrainment effects that start to occur during the later stages of the shock propagation ($> 1\ ms$, from the explosion). Hence, the exact value of the equivalent explosion energy cannot be obtained for the current experiments. This limitation can be circumvented by using the shock arrival time at a specific location (r). Using the experimental value of the Mach number ($M_s$) at a specific location as a reference, the theoretical value of $M_s$ is iteratively evaluated at that location using different values of $R_{o,}$ and the value of $R_o$ is obtained corresponding to a given experimental run. Thus, following Bach and Lee (1970) and Chandra et al. (2023), the theoretical shock arrival time ($t_{arr,Th}$) is evaluated by substituting the value of $R_o$ in Equation 3.11, which is in agreement with the experimentally obtained shock arrival time ($t_{arr,exp}$) at the droplet location (see supplementary figure S1).

Thus, the shock trajectories i.e., $R_s(t)$ and Mach number ($M_s$) evolution have been evaluated theoretically for the corresponding values of $R_o$ using j=2 for open-field blast wave (spherical assumption at the droplet location, $L_t + L_d >> w$). This has been plotted (solid lines) in figure 3i for the open-field blast wave cases (j=2), with origin (r=0) at the location of copper wire and time, t=0 at the time of the explosion. The experimentally measured shock trajectory evolution has also been plotted and is found to be in good agreement with the theory in the initial stages (~ 0 – 1 ms after the explosion). Similarly, the theoretical shock trajectories have also been obtained for the cylindrical blast waves (j=1) exiting the shock tube (focused cases), taking the reference time of the hypothetical explosion ($t_v = 0$) at the shock tube exit which is assumed to create a similar blast wave that expands cylindrically outward with the virtual origin at the shock tube exit (r=0). Similarly, the theoretical shock trajectory of the planar blast wave (j=0) traveling inside the shock tube from the time of explosion (t=0) of the copper wire has been obtained using the arrival time of the planar shock at the shock tube opening. The experimentally measured shock trajectories (points), and the theoretically obtained shock trajectories (solid lines) are plotted in figure 3f-h for different focused cases against the virtual time $t_v, in\ ms$ (where $t_v = 0$ is the instant the shock location is at the shock tube exit). The experimental shock trajectories are in really good agreement with the theoretical estimates (see figure 3f-h). From the plots, it can be clearly observed that the experimental shock trajectory initially follows the theoretical estimate of the planar shock trajectory (from the time of explosion) and then gradually shifts, deviates to follow the cylindrical shock trajectory estimate (with shock tube exit as the virtual origin). This establishes the previously hypothesized planar to cylindrical (centre at shock tube exit) transition of the propagating blast wave after exiting the shock tube opening.

The temporal variation of the shock location is tracked using high-speed Schlieren imaging and is plotted in figure 3k for all the cases. Figure 3j shows that all the plots merge into one single straight line in the log-log plot of shock radius ($R_s$) vs time ($t_v$) and the slope of the line is found to be near unity. It is shown that the scaling for the strong blast wave propagation is $R_s \sim {t_v}^{2/5}$ (exponent ~ 2/5) and for the weak acoustic limit, the weak blast wave propagation scaling is linear $R_s \sim$ t with (exponent ~ 1) (Díaz and Rigby 2022, Wei and Hargather 2021). This suggests that all the blast waves observed in the current experiments are near the weak blast wave acoustic limit (exponent near unity), however, the slopes of the individual cases (in log-log plot) are found to vary between 0.89 to 0.99. This shows that the current experiment is in the transition regime between the strong blast wave and acoustic limit. The plots in figure 3b-I shows the temporal variation of the Mach number ($M_s$) and the open-field blast wave cases show minimal variation in $M_s$ suggesting them to be near the weak blast wave limit, while the focused blast wave cases showed significant temporal variation of the $M_s$.

### 3.2. Interaction with a combusting droplet:

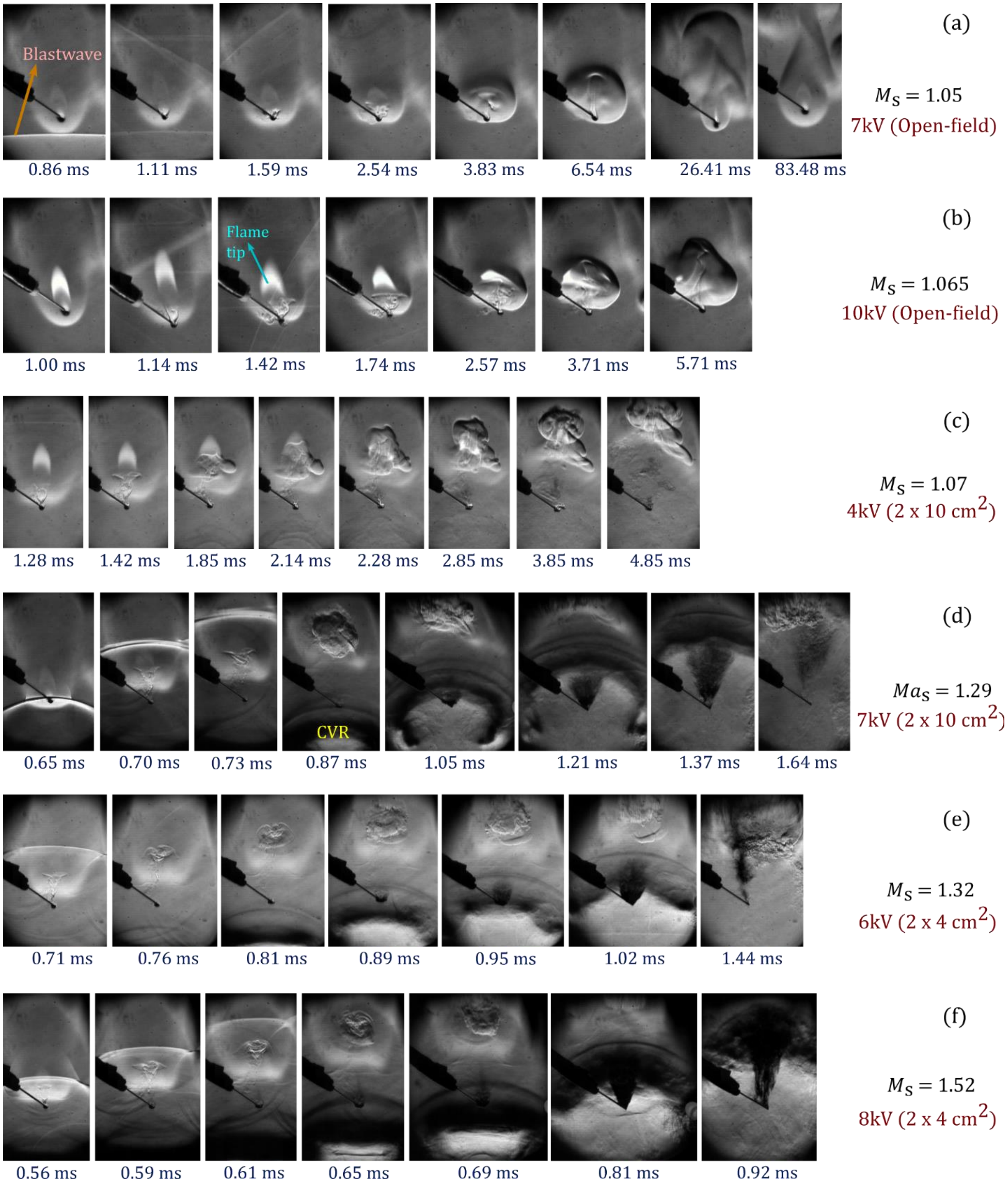

**Figure 4.** Time series of the Schlieren imaging depicting the overall interaction of the droplet flame with the shock flow for open-field configuration with charging voltage: (a) 7kV, (b) 10kV, for bigger shock tube channel (2 cm × 10 cm c/s) with charging voltages: (c) 4kV, (d) 7kV, and for the smaller shock tube channel (2 cm × 4 cm c/s) with charging voltages: (e) 6kV, (f) 8kV.

As mentioned before, the droplet is placed coaxially at a distance of $L_t + L_d$ from copper wire (which is $L_d$ from shock tube opening) in both Open-field and focused configurations, respectively. After the pendant droplet is ignited, the wire explosion is triggered to generate a blast wave. In the Open-field case, the blast wave expands radially outward towards the droplet, and in the focused case, the blast wave is directed along the shock tube, and it interacts with the droplet after it exits the shock tube. Figure 4 a,b shows the flame interaction with the blast wave generated in the Open-field configuration (no shock tube), whereas figure 4 c-f shows the flame interaction for focused blast wave configuration (with shock tube of the corresponding dimensions) for different charging voltages. Since $M_s$ varies temporally, the instantaneous value of $M_s$ when the blast wave passes the droplet location is considered as the reference value to characterize different cases and is mentioned in Figure 4, corresponding to each case. The subsequent discussion will categorize the different configurations of the shock tube at varying charging voltages as follows: 'B', 'S', 'Open' for big, small sections, and open-field configurations respectively, accompanied by their corresponding kilovolts (kV) values. For instance, the notation '4kV_Open' denotes a charging voltage of 4kV in an open-field configuration.

Along with the blast wave visualization, and density gradient contrast corresponding to the plume of hot gases around the flame, the bright flame tip is also visible in the Schlieren images. Figure 4 a,b depict the time series of the Schlieren images showing the blast wave propagation and flame response for Open-field configuration at two charging voltages. The shock Mach numbers ($M_s$) for the open-field configuration are less than 1.1 and the flame response is observed to occur over a longer time scale of order $\sim O(10^0 - 10^1)\,ms$. The flame is observed to liftoff in both the cases, however as shown in figure 4 a,b, total extinction occurs at higher charging voltage (10kV), and partial extinction followed by reignition is observed in case of 7kV and lower.

The time series presented in figure 4 c-f are the high-speed Schlieren snapshots during the interaction of the droplet flame with the focused blast wave (using the shock tube). Figure 4 c,d and Figure 4 e,f correspond to the focused cases using the shock tube with bigger (2 cm × 10 cm c/s) and smaller (2 cm × 4 cm c/s) cross-sectional dimensions, respectively. The higher $M_s$ obtained for the smaller section is higher compared to the larger section at the same charging voltage. Except for the 4kV charging voltage with big section, the Mach numbers obtained for all the focused cases are greater than 1.1. Furthermore, for similar $M_s$ values, the flame response behaviour for the focused cases is consistent with that of the open-field configuration. This can be observed in 4kV_B (big section) focused case with $M_s < 1.1$, where the flame responds over a longer time scales of order $\sim O(10^{-1} - 10^1)\,ms$ similar to Open-field configuration, see figure 4c. However, the flame response is within the period of the shock decay ($t \sim 0 - 1\,ms$, time from explosion) for higher $M_s$ (> 1.1). In all the focused cases figure 4c-f, a compressible vortex ring (CVR) is observed to form and interact with the droplet and flame after some delay. This interaction is observed to result in droplet breakup, which will be discussed in the later sections.

From the blast wave literature (Bach and Lee 1970, Goldstine and von Neuman 1963, Sedov 1957), it is known that the velocity profile at a given location imposed by a propagating blast wave decay temporally. The decay of the velocity profile behind the blast wave occurs till a time scale of $\sim 1\,ms$ (from the time of explosion), after which the velocity ($v_s$) decays down to zero and even to negative values (of lower magnitude). Interestingly, for open-field configuration (figure 4a,b), it can be observed that the time of the total flame interaction occurring between $\sim O(10^0 - 10^1)\,ms$ which is one order slower when compared to the shock decay time period of $t \sim (0.6 - 1)\ ms$ (from the time of explosion). During the time period of $t \sim 0.6 - 1\ ms$ corresponding to the decaying velocity profile imposed by the blast wave, the droplet flame is initially lifted off away from the droplet surface and it subsequently recedes towards the droplet. This lift-off is the consequence of the interaction of the blast wave profile with the droplet flame and subsequently flame starts to recede and reattach to the droplet as the decaying velocity profile ($v_s$) decays and approaches near zero near the droplet. However, even beyond this time period of influence of decay profile, $v_s$ ($t > 1\ ms$), see figure 4 a,b, where the velocity should have fully decayed to near zero, the flame is observed to exhibit a gradual lift off, albeit at a slower time scales of $\sim O(10^0 - 10^2)\ ms$. This can be attributed to a slower induced flow velocity ($v_{ind}$) at the time scales of $\sim O(10^0 - 10^2)\ ms$, which occur as a result of the entrainment effects at the edge of the expanding blast wave and thus deviating from the local velocity variation obtained from the theoretical formulation in section 3.1 (mass conservation is violated). The static pressure profile behind the blast wave (P) also decays temporally and decays below the ambient pressure locally at $t = t_d$ (see supplementary figure S2). These negative pressures (i.e., $P < P_{atm}$) behind the blast wave can be attributed to the occurrence of air entrainment effects. Thus, for the time $t < t_d$, the decaying profile of the blast wave is dominant and after $t > t_d$, the magnitude of the local velocity ($v_s$) becomes small and negative pressures occur at the droplet location drawing the bulk induced flow ($v_{ind}$) near the droplet.

However, the effects of $v_{ind}$ are only experienced at the droplet location after some delay beyond a certain time period (after $t = t_d$), depending on the velocity scale of the induced flow ($v_{ind}$). Based on the response time scales of the slower flame response for $t > t_d$, it can be concluded that the velocity scale of the induced flow ($v_{ind}$) is significantly slower compared to the shock flow. Thus, induced flow will affect the droplet flame only after it travels from the blast wave edge and reaches the droplet location. Thus, it can be concluded that the velocity decay profile ($v_s$) behind the blast wave is only valid during the initial stages of the interaction before the induced flow ($v_{ind}$) reaches the droplet location. This induced flow ($v_{ind}$) is responsible for the slower flame lift-off beyond $t > 1\ ms$, as shown in figure 4a,b. In all the cases, under the influence of the imposed flow, the fuel vapor plume itself is swept downstream (see figure 4) in response to either the velocity profile at the droplet due to the blast wave velocity profile ($v_s$) or the induced velocity scale ($v_{ind}$). This advection of the fuel vapor plume corresponds to the aforementioned flame lift-off phenomenon.

The entrainment effects (near the shock tube mounting at the cover plate) are also present in the case of shock tube focusing, and as this induced flow ($v_{ind}$) exits the shock tube, it encounters the ambient atmosphere. This results in the curling of the induced flow ($v_{ind}$) forming vortical structures as shown in figure 4c-f. These vortical structures show distinct contrast in the Schlieren images making them clearly visible (dark patches) and this can be attributed to the compressible nature of these vortical structures. Thus, these vortical structures will be referred to as compressible vortex rings (CVR) hereafter. The velocity scale ($v_{ind}$) of the CVR is also relatively slower compared to the shock flow which is evident from the CVR arrival time at the droplet. The time from the explosion is normalized using the pressure decay time below ambient at the droplet ($t = t_d$) given by $\tau = \frac{t}{t_d}$. Thus, the blast wave reaches and interacts with the droplet flame at $\tau = \tau_s$ and the static pressure at the droplet decays below ambient at $\tau = 1$.

Figure 5 shows the overall schematic of the different phenomena simultaneously occurring during the interaction of the blast wave and the droplet flame. As shown at the top of figure 5, The entire process of the interaction with the droplet flame occurs in two phases. The first stage is the blast wave – flame interaction corresponding to the time scale of $\tau_s < \tau < 1$ (approximately $t \sim 0.6 - 1\ ms$ after explosion). The second stage ($\tau > 1$) comprises of the induced flow – flame interaction ($t \sim O(10^0 - 10^1)\ ms$ after explosion) and simultaneous the induced flow CVR – droplet interaction ($t \sim O(10^0)\ ms$ after explosion). During the blast wave – flame interaction, the droplet flame responds directly to the propagating blast wave; thus, the droplet flame initially ($\tau < 1$) interacts with the decaying velocity profile ($v_s$) that is imposed by the blast wave (Bach and Lee 1970, Goldstine and von Neuman 1963). Later for $\tau > 1$, when the induced flow ($v_{ind}$) behind the blast wave reaches the test section, both the droplet as well as the droplet flame interact with the induced flow simultaneously. Since, the blast wave effects have fully decayed beyond $\tau > 1$, the flame response is only due to the induced flow ($v_{ind}$). The order of the magnitude of the induced flow ($v_{ind}$) can be reasonably assumed to be the same order as that of the lift-off speed of the flame base ($v_{b,lft}$) during this second stage of interaction.

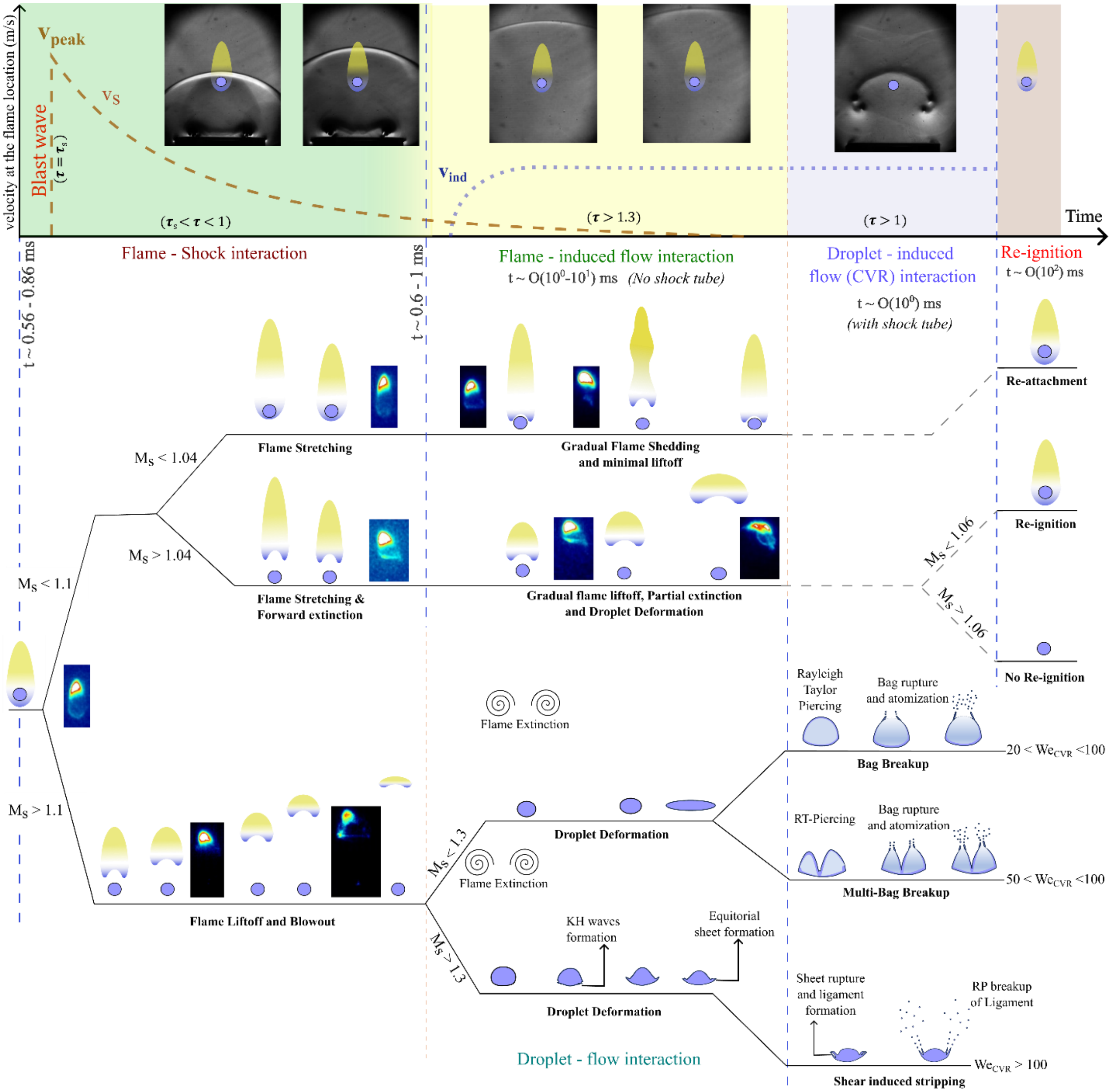

**Figure 5.** The schematic of temporal variation of the velocity during different stages of the interaction (top) plotted against normalized time, $\tau = \frac{t}{t_d}$. Schematic of the simultaneous response of the flame and droplet

to the blast wave profile ($v_s$) and induced flow ($v_{ind}$) at different stages of interaction. The coloured sub-figures are the instantaneous snapshot obtained from highspeed flame imaging.

It is to be noted that even though the induced flow – flame interaction regime and CVR – droplet interaction regimes are shown separately side by side in figure 5, these two interactions occur around a similar time scale ($t \sim O(10^0 - 10^1)\ ms$ after explosion). The green region in the plot shown at figure 5 (top) shows the trend of the temporal variation of velocity ($v_s$) at a given location as the blast wave propagates. When the blast wave propagates past the droplet, locally an instantaneous velocity peak ($v_{peak}$) is experienced due to the discontinuity imposed by the blast wave. Then the local velocity due to the blast wave profile ($v_s$) decays as the blast wave propagates further downstream (Bach and Lee 1970, von Neuman 1963). However, due to the entrainment effects, an induced flow is entrained behind the blast wave which reaches the droplet location after some delay during the later stages of the interaction i.e., $\tau > 1$ (i.e., $t \sim O(10^0 - 10^1)\ ms$ after the explosion). This induced flow ($v_{ind}$) is depicted in the indicative plot of the velocity variation in figure 5 (top). The green, yellow, and blue backgrounds in the plot represent the shock – flame interaction, induced flow – flame interaction (in the absence of shock tube), and droplet – induced flow CVR interaction (with shock tube) respectively. The approximate time scales corresponding to each of the interactions are mentioned in the figure. The schematic of the simultaneous response of both the flame and droplet during the different stages of interaction are depicted in figure 5.

Broadly the response of the flame and droplet to the imposed flow is divided into two regimes: low shock strengths ($M_s < 1.1$) and high shock strengths ($M_s > 1.1$), as shown in figure 5. Among the different experimental cases, all the open-field cases and the focused case, 4kV_B (charging voltage: 4kV, big section, 2 cm × 10 cm c/s), fall under the $M_s < 1.1$ behaviour. The rest of the higher shock strengths in the focused case follow the high shock strength ($M_s > 1.1$) behaviour. In the low shock strength regime ($M_s < 1.1$), the flame is observed to sustain beyond the initial blast wave – flame interaction stage ($\tau > 1$, figure 5, top). Whereas for $M_s > 1.1$ regime, the flame is observed to fully extinguish during the interaction with the decaying velocity profile ($v_s$) imposed by the blast wave ($\tau > 1$), i.e., green zone (figure 5, top). This is depicted under Flame – shock interaction regime in figure 5. Within the $M_s < 1.1$ regime, the flame behaviour can be further divided based on whether flame extinction occurs or not.

For $M_s < 1.04$, during the initial blast wave profile ($v_s$) interaction, $\tau < 1$ (green zone, figure 5), the flame lift-off is observed to be minimal (no forward extinction). However, for $M_s > 1.04$, the flame exhibits local extinction due to the critical strain rate at the forward stagnation point as a result of the externally imposed flow during this interaction stage. This is hereby referred to as 'forward extinction'. In both the cases for $M_s < 1.1$ regime, the flame does not extinguish during this initial stage. Subsequently, the flame starts to recede towards the droplet as soon as the velocity profile ($v_s$) at the flame decays and approaches zero around $t \sim 1\ ms$ (after the explosion), as depicted at the junction of the green and yellow regions in the plot shown in figure 4 (top). Subsequently, in the time period of $\tau > 1$ (yellow zone, figure 5), the slower induced flow ($v_{ind}$) reaches and interacts with the droplet flame, resulting in flame forward extinction and liftoff in $M_s < 1.1$ regime. In the case of $M_s < 1.04$, the induced flow velocity ($v_{ind}$) results in minimal lift-off ($h_{lft}$) of flame base accompanied by flame tip stretching and shedding, followed by reattachment of the flame onto the droplet surface and enveloping around the time scale of $\sim O(10^2)\ ms$ (depicted with brown background in figure 5). For $M_s > 1.04$, the flame tip shedding or stretching is not observed and the flame base undergoes slower but continuous lift-off achieving higher liftoff ($h_{lft} > 2d$) compared to $M_s < 1.04$ regime. Furthermore, for $1.04 < M_s < 1.06$, the continuous flame liftoff leads to partial extinction followed by subsequent reignition of the droplet around $t \sim O(10^2)\ ms$. However, for $M_s > 1.06$, full extinction without reignition is observed. All these sub-regimes are portrayed in the global schematic shown in figure 5. Consistency in flame behaviour based on Mach number ($M_s$) is observed, where the focused case 4kV_B ($M_s$ = 1.07) exhibited full extinction similar to 10kV_Open ($M_s$ = 1.065) as $M_s$ in both the cases fall under same sub-regime of $M_s$ > 1.06. Simultaneously, the droplet also interacts with the induced flow ($v_{ind}$) undergoing minimal deformation for open configuration but does not exhibit any breakup due to low values of $v_{ind}$.

For $M_s$ > 1.1, due to higher $M_s$ the velocity scale ($v_s$) associated with the decay profile behind the blast wave is higher. Thus, the flame lift-off is more dominant in this regime occurring at faster time scales of similar order as that of the shock decay time period ($t \sim 0.6 - 1\ ms$) for $\tau < 1$. Thus, the flame undergoes rapid continuous lift-off in response to $v_s$ imposed by the blast wave, ultimately leading to imminent extinction during the initial interaction stage i.e., $\tau < 1$ (green zone, figure 5). After extinction, the hot gases swept downstream and are observed to curl into a vortex ring (depicted in droplet – flow interaction of figure 5) and advect downstream due to the vorticity generation effect of the blast wave interaction due to RM instability (Picone and Boris 1988, Ju et al. 1988), as shown in figure 5c (rightmost image). Experimentally, it has been observed that the flame fully extinguishes before the induced flow CVR interacts with the droplet flame. Simultaneously, while the flame is interacting with the decaying profile ($v_s$) of the blast wave ($\tau < 1$), the droplet also interacts with the $v_s$ profile and subsequently with induced flow CVR ($v_{ind}$) undergoing either shear-induced stripping or Rayleigh-Taylor piercing. In all the cases, in response to the decaying velocity of the blast wave ($v_s$) for $\tau < 1$, the droplet exhibits continuous temporal deformation (increase in major axis length that is perpendicular to the flow direction). Additionally, perturbations on the windward surface of the droplet due to KH instability were observed for $M_s$ > 1.3 sub-regime due to decaying velocity ($v_s$). Later for $\tau > 1$, after some delay when the induced flow CVR ($v_{ind}$) reaches the droplet and interacts with it, shear-induced stripping is observed due to further growth of KH waves. The induced flow CVR ($v_{ind}$) is observed to reach and interact with the droplet quicker as the $M_s$ is increased. In the case of $M_s$ < 1.3 sub-regime, the KH waves are not formed due to lower velocity scales ($v_{ind}$), however, the droplet deformation continues leading to Rayleigh-Taylor piercing followed by bag breakup. This Rayleigh-Taylor piercing breakup mode is experimentally observed to be approximately one order slower compared to the shear-induced stripping mode. These observations will be detailed further in section 3.4. It is to be noted that, for the case of 4kV charging with big section ($M_s$ ~ 1.07), even though the flame follows the similar trend as shown in figure 5 for $M_s$ < 1.1 regime, due to the presence of the shock tube and focusing of the induced flow ($v_{ind}$) in the form of CVR, the CVR interaction with the droplet results in Rayleigh piercing breakup based on the Weber number of the CVR incident on the droplet.

#### 3.2.1.Shock – flame interaction ($\tau_s < \tau < 1$)

This section will primarily discuss the interaction of droplet flame for $\tau < 1$ (faster time scales) with the decay velocity profile ($v_s$) imposed by the blast wave (green zone, figure 5). For all the cases, the droplet flame is observed to immediately starts to respond to the propagating blast wave when it reaches the droplet location at $\tau = \tau_s$ (see figure 4). Figure 6 shows the temporal variation of the flame dimensions (in mm) during the interaction. The green background represents the first stage of interaction between the blast wave and the droplet flame. The orange data points represent the distance between the flame tip and the droplet, whereas the blue data points represent the distance between the flame base and the droplet (flame stand-off distance). This implies that the flame is vertically present in between the orange and blue data points with droplet location at the x-axis. The red dotted line depicts the approximate instantaneous location of the propagating blast wave. In figure 6, the shock strength increases from left to right due to the increase in charging voltage for the same shock tube configuration. Figure 6 a-b, figure c-d, figure e-f show the temporal flame dimension variation for Open-field, Big channel shock tube (2 cm × 10 cm c/s) and small channel shock tube (2 cm × 4 cm c/s) respectively.

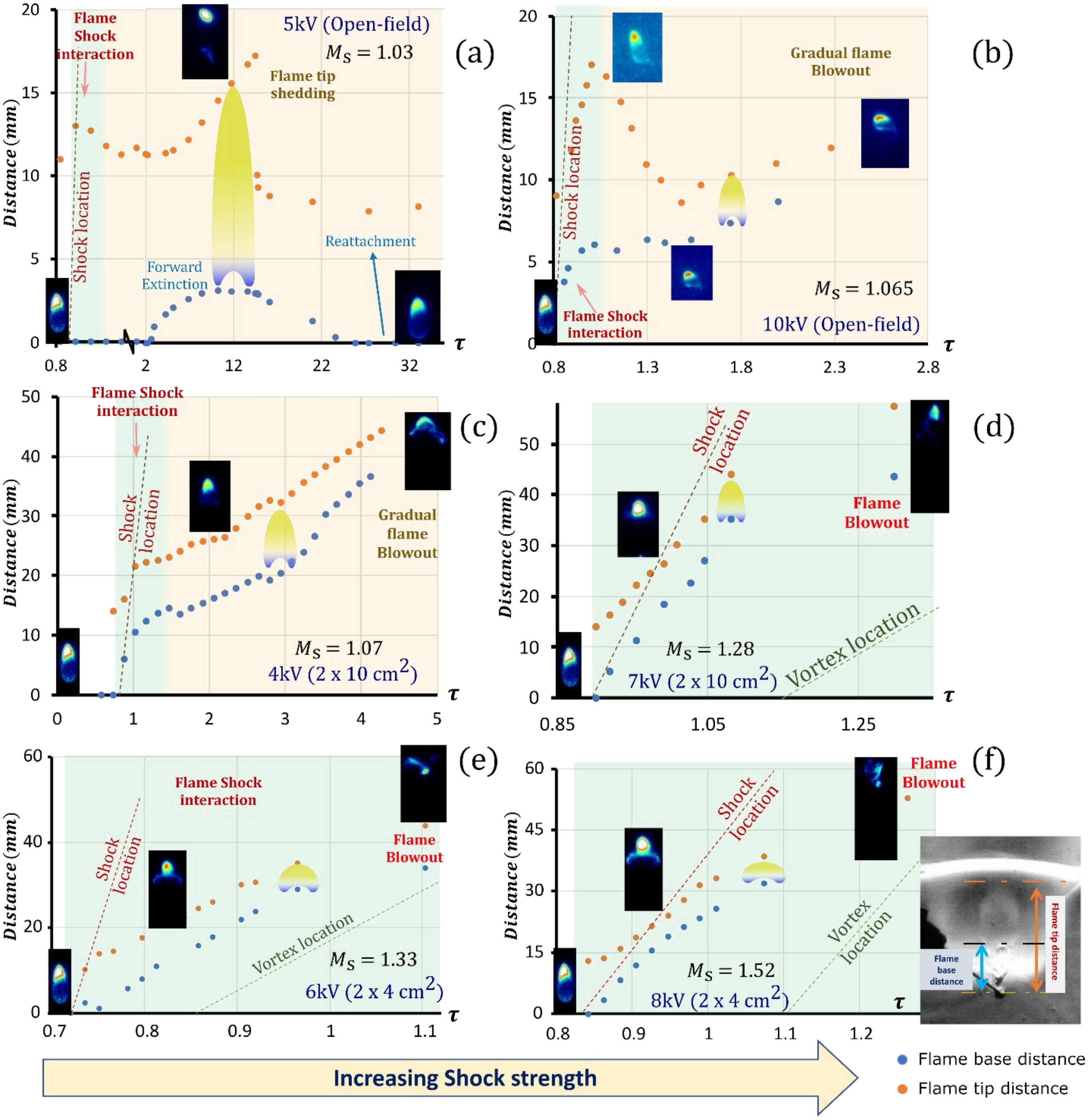

**Figure 6.** Temporal variation of the flame dimensions i.e., flame tip (orange) and base (blue) distances (mm) from the droplet surface during the initial interaction between blast velocity profile, $v_s$ (Green region) and later interaction between the induced flow, $v_{ind}$ (Yellow region) plotted for (a,b) open-field, (c,d) bigger channel (2 cm × 10 cm c/s), and (e,f) smaller channel (2 cm × 4 cm c/s) at different charging voltages. The flame dimensions are plotted against normalized time ($\tau$). The simultaneous shock and vortex location are shown in the plots with red and green dotted lines respectively and corresponding values $M_s$ were mentioned for each case.

During the first stage of shock – flame interaction ($\tau_s < \tau < 1$), in case of the open-field blast wave with $M_s$ (< 1.04) (figure 6a), the flame lift-off from the droplet surface remains minimal exhibiting flame tip stretching in the form of a quick jittery motion. As the blast wave passes by, the flame stretching is observed between $\tau_s < \tau < 1$ (almost immediately as the shock propagates) in response to the decaying velocity profile ($v_s$) imposed by the blast wave. No significant delay is observed between the flame response and the instant of shock interaction. Subsequently, the flame tip recedes towards the droplet (flame length reduces) as the velocity profile ($v_s$) decays at the flame, approaching near zero towards the end of the first stage interaction (green zone, figure 6). This is evident from the plot in figure 6a, which shows a local peak in the flame tip distance (orange) in the green region (shock – flame interaction phase) corresponding to the flame tip stretching. Later the drop in flame tip distance (orange) shows the receding of the flame tip distance after the shock passes. The plot also shows no significant flame lift-off (forward extinction) for this case (blue) in the green region. However, in case of a higher Mach number ($M_s$ ~1.065) (figure 6a), the flame undergoes local extinction at the forward stagnation point due to critical strain rate as a result of the externally imposed flow, followed by significant flame lift-off ($h_{lft}$ > 3d). Similar to the low $M_s$ case, the flame tip exhibits stretching followed by receding of the flame base and tip towards the droplet in response to the temporally decaying velocity profile ($v_s$) at the flame location. This flame tip stretching and receding is reflected in the plot in figure 6b in the form of a momentary spike in flame tip distance (orange) that is accompanied by the drastic increase in flame base lift-off (blue). The receding of the flame tip and base distance corresponds to the flame trying to reattach to the droplet after the blast velocity profile ($v_s$) decays significantly approaching zero around $\tau \sim 1$ (at the end of the green zone).

As mentioned before, higher confinement (focusing) of the blast wave results in higher $M_s$. The green zones in figure 6 c-d, figure e-f shows the plots of the temporal variation of the flame dimensions during $\tau_s < \tau < 1$ (shock – flame interaction phase) for bigger and smaller shock tube channels, respectively. The sub-figures are the instantaneous snapshots obtained from highspeed flame imaging. Since, the 4kV_B case ($M_s$ ~ 1.07) (bigger channel) has similar $M_s$ as that of 10kV_Open ($M_s$ ~ 1.065), similar flame dynamics is observed in both the cases. The flame shows forward extinction as well as significant lift-off ($h_{lft}$ > 3d) and flame tip stretching. The same can be seen in figure 6c which shows an initial rise in flame tip (orange) and flame base (blue) distances corresponding to flame stretching and flame lift-off, respectively, in the green region (shock – flame interaction phase, i.e., $\tau_s < \tau < 1$).

When the charging voltage is further increased in big channel (2 cm × 10 cm c/s), the flame lift-off during $\tau < 1$ (green zone) becomes more drastic as the higher $M_s$ (> 1.1) are achieved. The flame lift-off distance ($h_{lft}$) becomes more than five times the droplet diameter after interaction with the blast wave, leading to a blow-out. The flame lift-off is observed to continuously increases in response to the imposed $v_s$. The same is reflected in the plot in figure 6d where the flame base distance (blue) continuously increases drastically ($h_{lft}$ >10d) in the green region ($\tau_s < \tau < 1$) before the imminent blow-out. The rate of increase in flame tip and flame base distances is observed to increase as the shock strength is increased. This entire phenomenon is observed to occur directly In response to the imposed decaying velocity profile ($v_s$) at the flame ($\tau_s < \tau < 1$, green region). For shock focusing with smaller channel (2 cm × 4 cm c/s), higher $M_s$ are achieved at the same charging voltages ($M_s$ > 1.1). Thus, the flame dynamics observed are qualitatively similar to the $M_s$ > 1.1 cases of the bigger channel and the flame response is found to occur in even shorter time scales due to higher velocity scales ($v_s$), as seen in figure 6e,f. In both the focused cases, for $M_s$ > 1.1, full extinction of the flame is observed during this initial stage of interaction ($\tau_s < \tau < 1$).

#### 3.2.2. Theoretical estimate of the flame response to the decaying velocity profile ($v_s$)

The initial flame base lift-off (advection of the flame base in the downstream direction) in different cases in time period of $\tau_s < \tau < 1$ has been observed to be as a result of the interaction of velocity profile ($v_s$) with droplet flame. Thus, the temporal variation of $v_s$ at the droplet location needs to be evaluated during this stage.

Thus, from the blast wave formulation (section 3.1), Equation 3.10 can be used to obtain the velocity variation at a given location (r) at any given instant if the instantaneous shock radius ($R_s$) and shock Mach number ($M_s$) are known. Since it has already been established previously that the shock trajectory and $M_s$ are in good agreement with the theory (section 3.1 and figure 3f-i), the instantaneous values $R_s(t)$, $M_s(t)$ are measured experimentally, and the local velocity, $v_s(t)$, at the droplet location i.e., $r = L_t + L_d$ (for open-field case) or $r = L_d$ (for focused case) can be evaluated using Equation 3.10. The blast wave location is tracked along the centreline using Otsu's thresholding (in-built in ImageJ software) on the Schlieren images (section 2.1) to obtain the shock radius ($R_s$). Otsu's thresholding algorithm returns a single intensity threshold cutoff by maximizing the inter-class variance of the intensity values of a pixel which is used to convert the image into binary image to track the shock location. Thus, the instantaneous velocity variation ($v_s$) at the droplet flame location which is responsible for the advection of the flame base needs to be evaluated. The instantaneous flame base location ($h_{lft}$) has been extracted from the high-speed flame imaging using Otsu's thresholding and the advection velocity ($v_{b,lft}$) of the flame base is evaluated. The temporal variation of the local velocity at the droplet due to the blast wave profile ($v_{s,Th}$) based on theoretical shock trajectory (explained in section 3.1) is also evaluated.

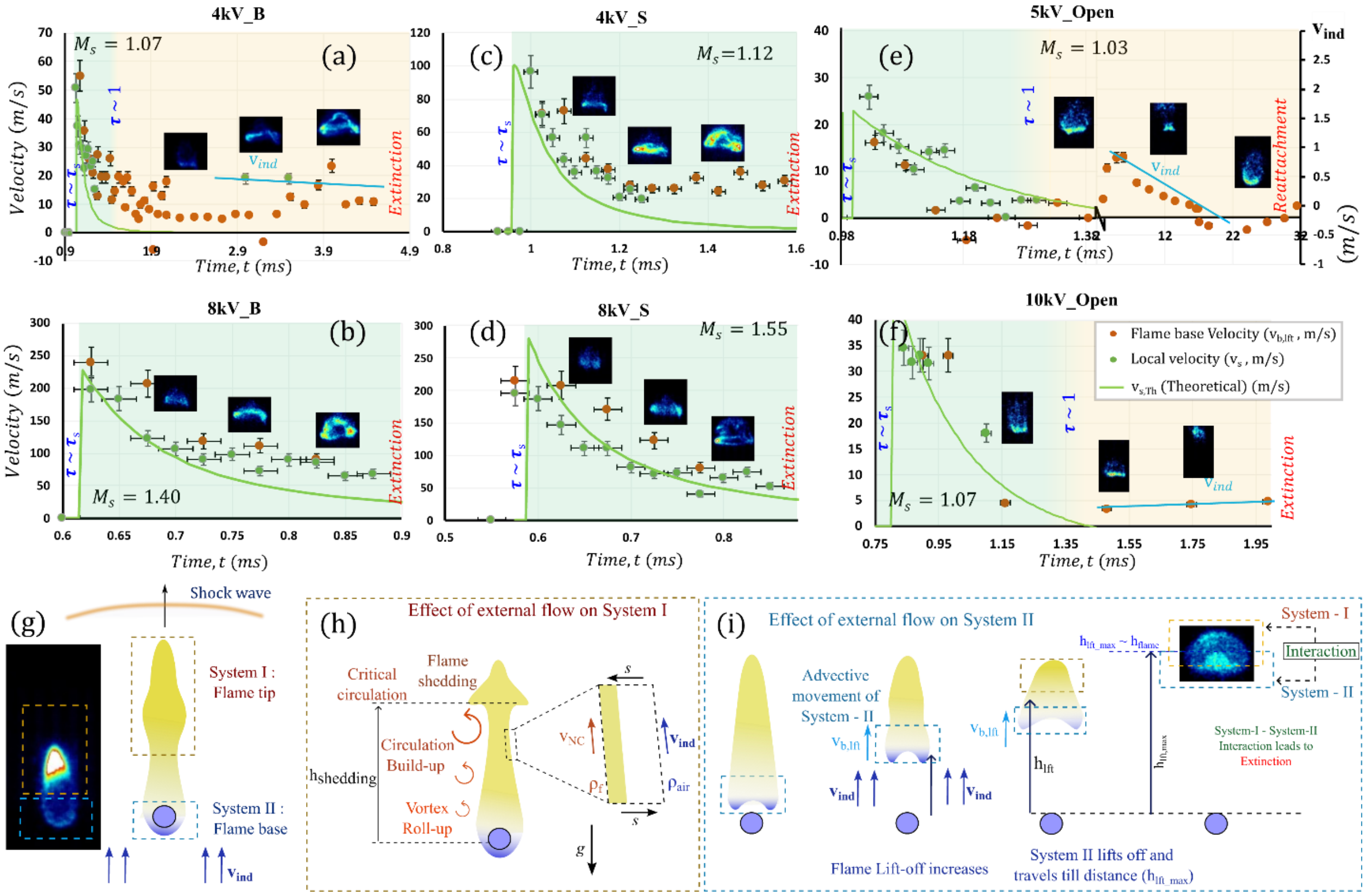

**Figure 7.** (a-f) Temporal variation of the local velocity ($v_s$) near the flame (based on instantaneous shock parameters), Temporal variation of the theoretically obtained local velocity near the flame location ($v_{s,Th}$) and Instantaneous flame base advection velocity ($v_{b,lft}$) plotted against time, t from the explosion for different cases 4kV_B, 8kV_B. 4kV_S, 8kV_S, 5kV_Open, 10kV_Open. The sub-figures are the flame snapshots of OH* Chemiluminescence depicting flame topology at a given time instant (for reference). (g) Schematic of droplet flame showing flame tip (system-I) and flame base (system-II), (h) Effect of externally imposed flow on flame tip (shedding mechanism), (i) Effect of externally imposed flow on flame base (extinction criteria).

Thus, the local velocity ($v_s$) calculated using instantaneous $R_s$ and $M_s$ (from experiments), the temporal variation of the local velocity based on theoretical shock trajectory ($v_{s,Th}$) and the instantaneous flame base advection (lift-off) velocity ($v_{b,lft}$) are plotted in figure 7a-f for different cases. As seen in figure 7a-f (green region), the flame base advection velocity ($v_{b,lft}$) is obtained to be in decent agreement with the decaying velocity profile ($v_s$) obtained from the blast wave formulation in the initial stages of the shock – flame interaction ($\tau_s < \tau < 1$). As the time progresses ($\tau > 1$), the entrainment effects lead to the deviation of the actual flow velocity (corresponding to the flame response) from the theoretical velocity profile ($v_s$) of the blast wave. Thus, as shown in figure 7b,c,d ($M_s > 1.1$), the flame undergoes continuous liftoff leading to full extinction in response to the imposed $v_s$ during the initial stage of shock – flame interaction (green region).

#### 3.2.3. Induced flow – flame interaction ($\tau > 1$)

As explained in previous sections, in all the cases, for $\tau > 1$, induced flow ($v_{ind}$) follows the propagating blast wave due to the entrainment effects. In the case of the Open-field blast wave experiments, a weak induced flow ($v_{ind}$) is observed behind the blast wave after a certain time delay (yellow region in figure 5). The induced air flow ($v_{ind}$) in case of open-field configuration cannot be distinctly visualized in Schlieren imaging, however, the effect of the induced airflow is clearly evident on the flame dynamics. After the initial stage of the interaction between the blast wave and the droplet flame, the flame is observed to show a gradual flame liftoff in response to the induced flow ($v_{ind}$) for $\tau > 1$. This interaction between the induced flow and the droplet flame is hereby referred to as the second stage of the interaction that follows the initial shock – flame interaction. For the Open-field cases, the blast wave reaches the droplet flame around t ~ 0.85 ms ($\tau = \tau_s$) whereas, the induced flow ($v_{ind}$) effects are observed in the flame response after ~1.5 ms i.e., the induced flow reaches the droplet flame location after some delay (i.e., $\tau > 1.3$). This shows that the induced flow velocity scales ($v_{ind}$) are significantly slower when compared to the shock flow, as it takes longer for the induced flow to reach the droplet.

For the lower $M_s$ (lower charging voltage) in open-field blast wave experiments (figure 4a), initially for $\tau_s < \tau < 1$, the flame tip stretches with minimal lift off and recedes during the initial interaction with the shock (see figure 6a,b green region). Later after a certain time delay as the relatively slower induced flow ($v_{ind}$) reaches the droplet flame (around $\tau > 1.3$), the flame again starts to lift off in response to the induced flow ($v_{ind}$) at slower time scales after 2 ms (see figure 6a yellow region). This lift-off is accompanied by a flame shedding event for $M_s < 1.04$ sub-regime, and the flame lifts off to greater heights ($h_{lft} > 3d$) as $M_s$ is increased due to higher $v_{ind}$. At lower $M_s$, this lifted flame reattaches to the droplet and eventually attains an enveloped state similar to pendant droplet flame prior to the interaction for $M_s < 1.04$. Furthermore, partial extinction ($h_{lft} > 3d$) and re-ignition are observed for $1.04 < M_s < 1.06$.

On the contrary, for higher $M_s$ in open-field cases ($M_s > 1.06$, see figure 4b), when the induced flow ($v_{ind}$) reaches the droplet flame for $\tau > 1$ (figure 6b yellow region), the lifted flame exhibits liftoff similar to the lower $M_s$ case, however, the gradual liftoff occurs continuously during the interaction with the induced flow ($v_{ind}$) and leads to an imminent blowout. This is also observed in the temporal variation of the flame base (blue) in figure 6b (yellow region) where the flame base distance from the droplet increases further during the induced flow interaction phase, which finally results in blowout. This blowout during the interaction with the induced flow has been referred to as "Gradual Blowout" in figure 6b as this extinction occurs after some time delay after the blast wave interaction.

All these events during induced flow interaction phase ($\tau > 1$) in open-field cases, are significantly slower (occurring at t ~ 2 – 30 ms) compared to the initial interaction with the shock i.e., $\tau_s < \tau < 1$ (green region). While the timescale of the initial shock – flame interaction is of the order of O ~ $10^{-1}$ ms, the induced flow

– flame interaction occurs at a time scale of order O ~ $10^0$ ms, which is slower because of the lower velocity scales associated with the induced flow ($v_{ind}$) compared to the flow behind the shock.

The induced flow velocity ($v_{ind}$) is depicted in figure 7a,e,f using light-blue solid line denoted by '$v_{ind}$' for the corresponding runs. In these three cases i.e., 4kV_B ($M_s$ ~ 1.07), 5kV_Open ($M_s$ ~ 1.03) and 10kV_Open ($M_s$ ~ 1.065): the flame is sustained beyond the initial interaction with the blast wave decay profile (green region) and subsequently, it interacts with the induced flow (yellow region). Since, the induced flow ($v_{ind}$) cannot be experimentally obtained for the open-field cases, and the flame is only responding to the induced flow in yellow region, the flame base advection velocity ($v_{b,lft}$) is assumed to be of the same order as that of the induced flow ($v_{ind}$). In case of 5kV_Open (see figure 7e), the flame base undergoes gradual lift-off accompanied by flame tip shedding in response to this induced flow ($v_{ind}$) and then subsequently reattaches to the droplet around t ~ O($10^2$ ms). However, in the 8kV_Open case, the flame undergoes gradual lift-off leading to partial extinction in response to $v_{ind}$, and subsequently reignites and reattaches to the droplet after a time period of t ~ O($10^2$ ms). Furthermore, the 10kV Open-field case undergoes full extinction in response to $v_{ind}$ without reignition and reattachment. Both 8kV and 10kV open-field cases do not show flame-shedding phenomena. It is to be noted that for focused cases ($M_s$ > 1.1), the flame extinction is observed to occur during the initial blast wave interaction (green region), and the flame does not survive to interact with the incoming induced flow ($v_{ind}$).

#### 3.2.4. Shedding criteria of the flame

Experimentally it has been observed that for the current droplet flame considered, the flame shedding only occurs at the low shock Mach number ($M_s$ < 1.04) for $\tau > 1$ during the interaction with $v_{ind}$, which corresponds to 5kV_Open case (no shock tube). The shedding events are observed in the time scale of order t ~ O($10^1$ ms) where the velocity scale due to the blast wave profile ($v_s$) has completely decayed, and the induced flow ($v_{ind}$ ~ O($10^1$) m/s) is the dominant velocity scale present during the shedding phenomena. Thus, compressible effects are minimal corresponding to the low velocity scales of $v_{ind}$ in this time scale of $\tau > 1$, where flame shedding is observed. The schematic of the flame shedding mechanism is depicted in figure 7h (left), where due to the buoyancy-induced instability, the hot gases around the flame accelerate along the flame length near the shear layer leading to continuous vortex rollup. This leads to a continuous feeding of circulation from the locally perturbed region near the droplet (fuel source) along the flame length which eventually reaches a critical value leading to flame puffing or shedding. As shown by Xia and Zhang (2018), in the context of diffusion flames (Ri→ ∞), the gravitational (buoyancy) term is pivotal for the vorticity generation. Thus, the vorticity transport equation considered near the flame tip (system-I) can be written as follows:

$$\frac{D\omega}{Dt} = \frac{\rho_a}{\rho^2}(\nabla\rho \times g) + \nu\nabla^2\omega \tag{3.12}$$

$\omega$ is the vorticity, $\rho$ is the local density at a given location, $\rho_a$ is the air density and $\nu$ is fluid viscosity. The last term on right-hand side (R.H.S.) of the above equation represents the vorticity diffusion term can be neglected when compared to the first term (baroclinic term due to buoyancy) (Xia and Zhang 2018). Proceeding in a similar approach as provided by Xia and Zhang (2018), the temporal variation of the circulation in the control volume enclosing the shear boundary as shown in figure 7h (right) is given by the summation of the initial circulation strength fed due to the local velocities on either side of the shear layer and the buoyancy induced vorticity buildup, shown below:

$$\frac{d\Gamma}{dt} = \rho_a g\left(\frac{1}{\rho_a} - \frac{1}{\rho_f}\right)\Delta h + \frac{d\Gamma_{initial}}{dt} \tag{3.13}$$

In the above equation, $\rho_f$ represents the density of the hot gases inside the flame, $\Delta h$ represents the length scale associated with the control volume, $\Gamma$ represents circulation in the control volume at a given height and $\Gamma_{initial}$ represents the initial circulation present in the system due to the local velocity. Hence, for the droplet interacting with flow imposed by the blast wave in current experiments, $d\Gamma_{initial} = (-v_{NC} + v_{ind})dh$ as shown in figure 7h, because $v_{ind}$ is the dominant velocity scale imposed. Since $v_{ind}$ has been observed to be the dominant convective velocity experimentally, dh can be scaled as: $dh \sim (v_{ind} + v_{NC})dt$. Thus, $d\Gamma_{initial} = (-v_{NC} + v_{ind})^2 dt$. Here, $v_{ind}$ is the induced flow velocity which can be experimentally obtained.

Thus, the circulation buildup equation becomes:

$$\frac{d\Gamma}{dt} = \rho_a g\left(\frac{1}{\rho_a} - \frac{1}{\rho_f}\right)\Delta h + {v_{ind}}^2 - {v_{NC}}^2 \tag{3.14}$$

Thus, the rate of circulation buildup depends on the buoyancy-induced instability and the externally imposed flow ($v_{ind}$). The magnitude of $v_{ind}$ for 5kV_Open case is of similar order as the velocity scale observed by Pandey et al. (2021). Thus, using the scaling for buoyant flickering i.e., $t_{shd} \sim \sqrt{h/g}$ (Pandey et al. 2021), substituting in the circulation buildup equation and integrating the equation 3.14 on both sides over one shedding cycle (where the critical circulation is reached for shedding to ensue):

$$\Gamma_{critical} = \frac{\rho_a g^{1/2} k}{3}\left(\frac{1}{\rho_a} - \frac{1}{\rho_f}\right){h_{sh}}^{3/2} + k({v_{ind}}^2 - {v_{NC}}^2)g^{-1/2}{h_{sh}}^{1/2} \tag{3.15}$$

Thus, flame shedding occurs at a height of $h_{sh}$ where the circulation buildup will reach the critical value ($\Gamma_{critical}$) and the time scale of this shedding is $t_{shd}$. Experimentally, for 5kV_Open, the time scale of the flame shedding is observed to be around ~ 15 ms which is of the order of ~ $O(10^1)$ ms during the interaction with $v_{ind}$ ($\tau > 1$). This is evident from figure 6a spike in the flame tip distance (orange) in the yellow region. However, in all the other open-field cases where the flame extinction is observed, the time scale of extinction is found to be of the order of ~ $O(10^0)$ ms which is one order faster compared to the flame shedding time scale. Thus, there is no sufficient time for the circulation buildup to reach critical value (that ensues shedding) before the system-II (flame base) reaches the system-I (flame tip) resulting in and extinction event due to higher advection velocity of system-II at higher $M_s$.

### 3.2.5. Extinction criteria of the flame

Figure 8a shows the temporal variation of the net local velocity ($v_{comb}$) estimated at the droplet location as the blast wave propagates radially outward, plotted for different cases, where, $v_{comb} = v_s + v_{ind}$. The initial spike in velocity corresponds to the velocity imposed at the droplet ($v_s$) between $\tau_s < \tau < 1$ and the secondary spike/discontinuity in the velocity corresponds to the arrival of the induced flow vortex having a velocity scale of $v_{ind}$ (obtained experimentally) that arrives at the droplet after some delay ($\tau > 1$). The droplet flame response to the externally imposed flow is observed to manifest at the flame base (lift off) and flame tip (shedding) independently. Thus, as shown in figure 7g, the droplet flame tip and flame base are considered to be two independent systems: System-I (flame tip) and System-II (flame base). Figure 7i depicts the advection of the flame base (system-II) towards system-I. This advection (lift-off) of system-II is dependent on the external flow imposed on the droplet flame. The solid green line figure 7a-f shows the theoretically obtained temporal variation of the decaying local velocity ($v_{s,Th}$) imposed by the blast wave at the droplet flame location. The flame lift-off ($h_{lft}$) is influenced by only the blast wave profile for $M_s > 1.1$, however, for $M_s < 1.1$, the induced flow ($v_{ind}$) also affects the flame base advection velocity ($v_{b,lft}$). Thus, the combined velocity ($v_{comb}$) of decay profile at the droplet flame location ($v_{s,Th}$) and the subsequent induced flow velocity ($v_{ind}$) determines the overall advection of the flame base (system-II).

$$\mathrm{v}_{comb}(t) = \mathrm{v}_{s,Th}(t) + \mathrm{v}_{ind,inst}(t) \tag{3.16}$$

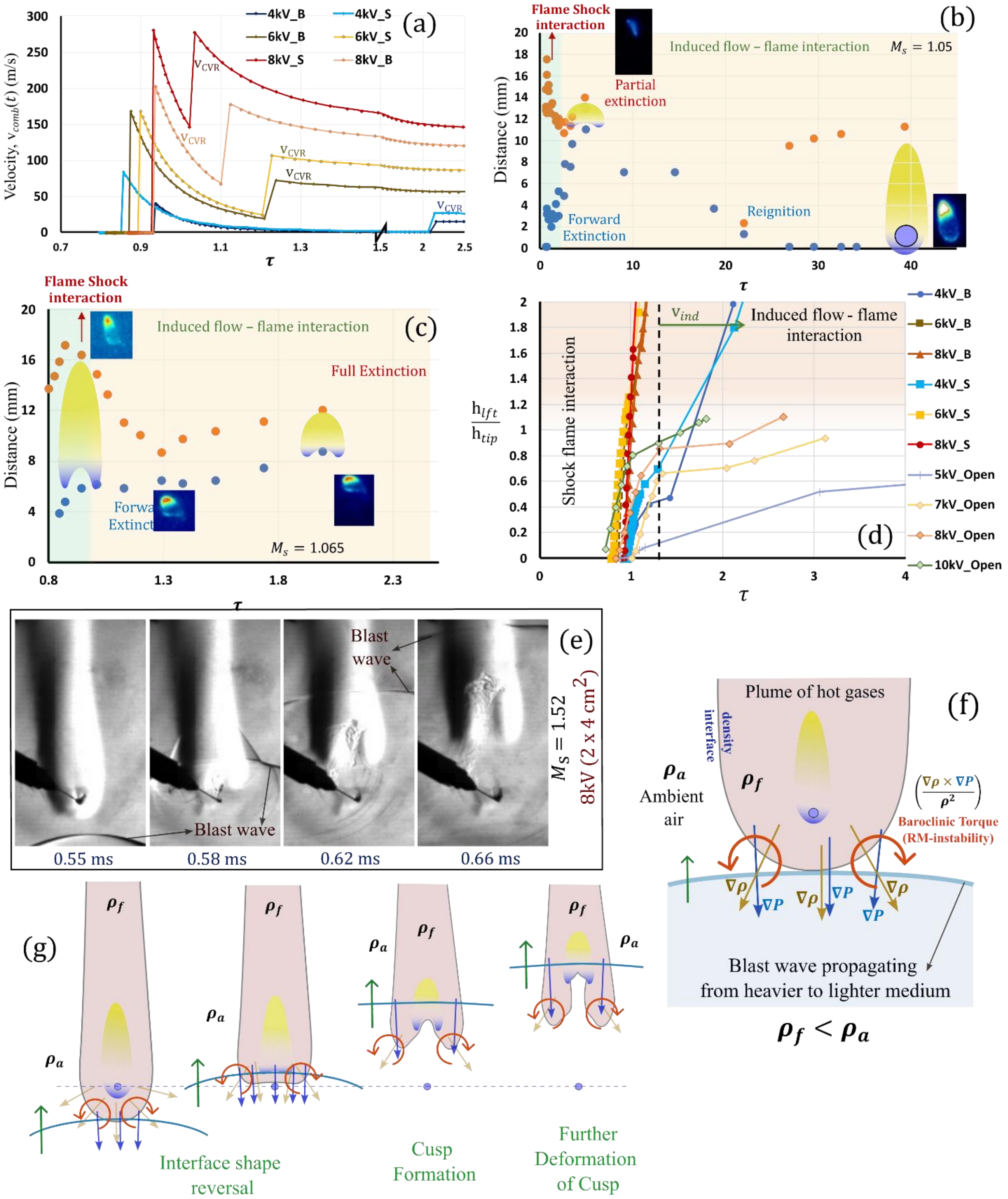

**Figure 8.** (a) Temporal variation of the total velocity at the droplet location that is obtained by the summation of the instantaneous velocity estimated using the blast wave formulation and the induced flow; (b,c) Temporal variation of the flame tip and base dimensions depicting the partial extinction, reignition for 7kV_Open, and full extinction for 10kV_Open. The sub-figures are the instantaneous snapshot obtained

from highspeed flame imaging; (d) *Extinction criteria:* lift-off ratio vs time for different cases. (e) Time series snapshots of Schlieren flow visualization showing the shape deformation of the windward density interface of the plume of hot gases around the droplet flame during interaction with shock; (f) Schematic of blast wave and the windward density interface of the hot gas plume around the droplet flame depicting the density and imposed pressure gradient directions that governs the baroclinic torque: Richtmyer-Meshkov instability (RMI); (g) Schematic of the shape deformation phenomenon of the windward density interface of the hot gas plume during interaction with the blast wave.

Since it has been established that the flame base liftoff velocity is a direct response to the local velocity at the flame (see figure 7a-f), The cumulative area under the plot $v_{comb}(t)$ vs time gives the total distance travelled by the flame base ($h_{lft,max}$). Experimentally, it has been observed that the flame extinction occurs when the advection velocity of system-II (flame base) is sufficient to traverse the flame length and reach the system-I, and thus, interact with it. This interaction between system-II and system-I results in a critical strain rate leading to flame extinction. Thus, the advection of system-II towards downstream reaching system-I (flame tip) is the criteria for extinction. Thus, the ratio of $h_{lft}$ and $h_{tip}$ (flame tip distance) is plotted against time in figure 8d. Figure 8d shows that the lift-off ratio ($h_{lft}/h_{tip}$) gradually increases with time as the flame interacts with the total local velocity imposed by the external flow. The vertical dotted line in figure 8d represents the approximate time instant when the velocity profile behind the blast wave decays and approaches near zero ($\tau > 1$). The left side of the vertical dotted line represents the initial shock – flame interaction ($\tau_s < \tau < 1$) and the right side of the dotted line represents the induced flow – flame interaction ($\tau > 1$). The horizontal dotted line corresponds to the lift-off ratio approaching unity. It can be observed that for focused blast wave cases at higher $M_s > 1.1$, the lift-off ratio ($h_{lft}/h_{tip}$) crosses unity during the initial interaction with the decaying velocity profile. This implies that for these cases, system-II interacts with system-I during the initial shock – flame interaction ($\tau_s < \tau < 1$) which is in agreement with the experiments.

On the contrary, for lower $M_s < 1.1$, the lift-off ratio ($h_{lft}/h_{tip}$) does not approach unity during the initial interaction ($\tau_s < \tau < 1$). That means, the flame is sustained beyond $\tau \sim 1$, and it subsequently starts to interact with the induced flow ($v_{ind}$) for $\tau > 1$. During the induced flow – flame interaction ($\tau > 1$), the lift-off ratio ($h_{lft}/h_{tip}$) is observed to reach unity for $M_s > 1.04$ (except 5kV_Open). This suggests that the flame extinction occurs during the interaction with the induced flow ($v_{ind}$) for lower Mach number cases except for 5kV_open case for which the lift-off ratio does not reach unity. This is consistent with the experimental observation of the absence of extinction occurrence in 5kV_Open. Furthermore, the lift-off ratio of 10kV_open ($M_s \sim 1.065$) approached unity with a steeper slope compared to 7kV_open ($M_s \sim 1.05$), 8kV_open ($M_s \sim 1.055$). This suggests that even though the extinction occurs in all the cases, the velocity ($v_{b,lft}$) of the flame base (system-II) advection is faster in case of 10kV_open compared to 7kV_open and 8kV_open. This higher velocity of system-II advection can be attributed to the occurrence of full extinction (due to critical strain rate) in the case of 10kV_open (see figure 8c). Additionally, the shallow slope of 7kV_open and 8kV_open indicates a lower advection velocity of system-II resulting in lower strain rates during its interaction with system-I (when the lift-off ratio reaches unity). This might be attributed to the occurrence of partial extinction in these two cases (7kV_Open, 8kV_Open), which resulted in reignition and reattachment of the flame subsequently (see figure 8b).

### 3.2.6. RM instability during flame – shock interaction:

Figure 8e shows the Schlieren snapshot time series of the response of the hot plume (around the flame) during the interaction with the blast wave ($\tau_s < \tau < 1$) for the case of 8kV_S. It is evident from the images that during the initial blast wave interaction, the hot plume is swept downstream, which corresponds to the flame base lift-off event. Along with this, it can also be observed that the topology of the hot plume and the flame alter during the interaction with the blast wave. As shown in figure 7a-d (sub-figures), the flame is observed to deform forming a vortical structure for ($M_s > 1.1$) during the initial stage of interaction with the blast wave which can be attributed to the vorticity generation in the flame region as a result of shock propagation (Picone and Boris 1988, Ju et al. 1988). The vorticity is generated in the flame region due to Richtmyer – Meshkov instability (RMI; a special case of Rayleigh – Taylor instability) that occurs when the local pressure gradient (imposed by the shock) being misaligned with the density gradient across the plume interface. The pressure gradient is caused by the acceleration field (gravity) in RT instability, whereas, in case of RM instability, the pressure gradient is caused by the propagating shock wave (Zhou et al. 2021). Considering the vorticity transport equation at the windward side of the interface of the hot plume around the flame (see figure 8f):

$$\frac{D\omega}{Dt} = u.\nabla\omega - \omega.\nabla\mathrm{u} + \frac{\nabla\rho\times\nabla\mathrm{P}}{\rho^2} + \nu\nabla^2\omega \tag{3.17}$$

The third term on the right-hand side (R.H.S.) represents the baroclinic component of the vorticity transport equation that occurs due to the misalignment of the density and pressure gradients which predominantly contributes to the vorticity generation. Furthermore, this baroclinic term is shown to be most pertinent to the immediate discussion of RM instability by Zhou et al. 2021 compared to the first term (vortex stretching), second term (vorticity dilation) and the last term (vorticity dissipation) on the R.H.S. even during shock wave interaction. Thus, this baroclinic vorticity term is activated when pressure and density gradients were misaligned which deposits vorticity on the density interface upon passage of the shock wave. This vorticity causes the perturbations on the density interface to deform, which leads to RM instability perturbation growth. Thus, depending on the curvature and geometry of the flame region (blob of density inhomogeneity) the baroclinic vorticity is generated in the system which eventually roll up into vortex filaments or vortex rings (Picone and Boris 1988), as shown in figure 7a-d (sub-figures). As consequence of this RMI-induced vorticity generation, the plume around the flame also exhibits shape deformation during its interaction with the blast wave.

The general behaviour of RM instability perturbation growth at an interface strongly depends on Atwood number (A), which signifies the difference in density of the two mediums on either side of an interface and it is given by:

$$A = \frac{\rho_2 - \rho_1}{\rho_2 + \rho_1} \tag{3.18}$$

Where, $\rho_1$ and $\rho_2$ are the densities of the medium on either side of the interface and the shock wave is considered to propagate from medium 1 ($\rho_1$) to medium 2 ($\rho_2$). If $A > 0$, then the interface perturbation spikes simply grow towards the lower-density material. However, for $A < 0$, phase inversion is observed where the perturbations begin to deform, forming inverted spikes that grow in opposite direction into the low-density side (Sterbentz et al. 2022).

When the shock propagates downstream across the interface, it interacts with the flame region containing a lower density hot plume (around the flame). In current experiments, the shock enters from a denser medium (air) to a rare medium (plume of hot gases), thus the Atwood number $A < 0$, which leads to phase inversion of the perturbations. It is also reflected in the experiments as shown in figure 8e where, the shape of the windward density interface of the hot plume is initially convex upstream (downwards) inverts exhibiting a concavity (facing upstream) after the initial interaction with the blast wave (see figure 4c-f and supplementary figure S3). This phenomenon of the reversal of the interface curvature is due to Richtmyer-Meshkov instability which occurs when the shock wave interacts with a density interface propagating from denser (unburnt gases) to rarer (burnt gases) medium, i.e., $A < 0$ (Yang et al. 2023, Sterbentz et al. 2022,

La Flechea et al. 2018). The schematic of this instability is depicted in figure 8g showing the phase inversion of the curvature of the density interface leading to the formation of a cusp-like structure. This cusp-like structure further grows into the lighter fluid side and subsequently curling-up forming vortical structures near the flame base. This leads to the aforementioned alteration in the flame topology forming vortical structures (as shown in figure 7a-d, sub-figures) as the consequence of the RMI-induced vorticity generation.

Kramer et al. (2010) investigated the single-mode RMI and showed that for weak and intermediate shocks, the impulsive model of the Rayleigh-Taylor instability is adequate for the prediction of the perturbation growth rate. Hence, this is applicable for the current experiments where the blast wave is near the weak shock limit. The amplitude growth due to Rayleigh-Taylor instability (a) of the single-mode perturbation on a discontinuous interface is given by (Zhou 2017, Zhou et al. 2021):

$$\frac{d^2a}{dt^2} = gkAa \qquad (3.19)$$

Where, $k = 2\pi/\lambda$ is the wavenumber, 'g' is acceleration and 'a' is perturbation amplitude.

Following by Zhou 2017, Zhou et al. 2021, for a varying acceleration i.e., $g(t) = \Delta u\, \delta(t)$, (using Dirac delta function) the initial impulse corresponding to the velocity jump imparted by the shock is given by $\int g(t)dt = \Delta u$, which can be integrated to obtain the following equation for perturbation growth during shock interaction.

$$\frac{da}{dt} = kA\Delta u a_o \qquad (3.20)$$

Where, '$a_o$' is the initial amplitude of the perturbation and '$\Delta u$' is the velocity jump imposed by the blast wave i.e., (~ $v_{peak}$), and 'A' is post-shock condition Atwood number across the density interface.

### 3.3. Effect of shock focusing using a shock tube

Unlike the open-field case, in focused cases (with shock tube) the induced flow behind the blast wave has to exit from the shock tube into the surrounding ambient. This causes the flow to curl as it exits the shock tube forming vortical structures. Researchers like Qin et al. (2020), Zare-Behtash et al. (2008, 2009), Zhang et al. (2014), Ahmad et al. (2020) investigated the compressible vortex rings (CVR) exiting as shock tube. The formation dynamics of CVR have been investigated and different types of CVRs: shock-free, with embedded shock, and with secondary vortices. It has been shown that the CVRs roll up and grow in size and pinch off from the shock tube exit as they propagate downstream. In current experiments, the CVR is a consequence of a blast wave generation (using wire-explosion technique) due to which the current CVR observed needs to be characterized experimentally. From the experiments, it is observed that, at lower $M_s$ numbers (lower charging voltage), the vortex formed tends to dissipate quickly as it propagates downstream. The induced flow vortex has to travel $L_d$ ~ 35 mm outside the shock tube in order to interact with the droplet flame. This induced flow vortex is visually noticeable in the experimental images in figure 4c-f. Similar to the blast wave, the induced flow vortex is observed to travel faster in the case of the smaller shock tube channel (2 cm × 4 cm c/s) compared to the bigger channel (2 cm × 10 cm c/s). It is to be noted that, unlike the literature, the CVR behind the blast wave in current experiments did not exhibit any shock-cell structures in its trailing wake that are observed behind shock waves in the literature (Ahmad et al. 2020). This behaviour qualitatively matches with the observations of experiments by Chan et al. (2016) where the distinct shock cell structures are observed in CVR wake in case of shock tube exhaust of the compressed-air driven shock wave and no visible shock cell structures are observed in CVR wake in case of shock tube exhaust of the explosively driven shock wave.

As explained before, for low shock Mach numbers, $M_s < 1.1$, the flame sustains after its initial interaction with the velocity profile of the blast wave (~ 0.8 – 1.5 ms) during which the flame undergoes lift-off as well as stretching. For the 4 kV_B case i.e., ($M_s$ ~ 1.07), the flame follows a similar trend as that of the 10kV_Open case, where the flame survives through the initial shock interaction and then starts to interact with the induced flow ($v_{ind}$) when it reaches the droplet location after some delay. During this interaction (~ 1.5 – 5 ms), the flame lift-off continuously increases significantly (>10d) as the flame base is advected downstream due to $v_{ind}$ before the imminent blowout (see figure 4c).

As the charging voltage is increased in case of bigger channel ($M_s > 1.1$), the induced flow vortex behind the blast wave becomes more distinct as shown in figure 4d. However, by the time the induced flow reaches the droplet, the flame blowout would have already occurred during the initial shock – flame interaction phase due to higher $M_s$. Similar blowout during initial shock – flame interaction is observed in case of smaller channel as well, due to the higher focusing effect resulting in higher $M_s$ ($M_s > 1.1$). This can be seen in figure 6d and figure 6c where the flame extinction has already occurred in the green region (shock – flame interaction) when the flame base (blue) is advected downstream rapidly to interact with the flame tip (orange). In all the cases, as the flame extinguishes, the hot gases at the flame rise up, they curl to form a toroidal structure due to Rayliegh-Taylor instability (see figure 4).

### 3.3.1. Induced flow vortex characterization:

As explained before, the induced flow begins to curl as it exits the shock tube channel forming a vortical structure. The vortical structures are visibly noticeable in the Schlieren imaging suggesting density variation in the vortical flow behind the blast wave. Thus, it can be concluded that it is a compressible vortex. Figure 9a shows the schematic of the flow exiting the open end of the shock tube. It can be seen in figure 9a that the blast wave first exits the shock tube followed by the induced flow that travels at lower velocity scales compared to the blast wave. As discussed before, the blast wave has a decaying velocity profile behind it which is depicted in the schematic. The propagating blast wave imposes a velocity jump ($v_s$) or discontinuity (compared to the ambient downstream of it), and the velocity monotonically decays in amplitude behind the blast wave till the induced flow vortex arrives at the location.

Figure 9b shows the time series of the flow exiting the shock tube for low shock strength i.e., 4kV_B case ($M_s$ ~ 1.07). The Shock exiting the shock tube initially is planar near the exit. However, as it propagates downstream, the blast wave attains curvature as it expands radially outward away from the shock tube. This is the planar to cylindrical transition of the blast wave depicted in figure 3a which has been explained in detail in section 3.1. The induced flow follows the blast wave and exits the shock tube with some delay, which is evident from the slow-moving vortex at the shock tube exit. The vortex ring is not strong enough and its translation velocity is significantly slower compared to the blast wave for 4kV_B case. Furthermore, the vortex starts to dissipate before reaching the droplet (see figure 4c.).

However, in the case of higher charging voltages with the bigger channel i.e., 8kV_B case (Figure 9c), the Mach number is higher, and the blast wave shows higher contrast indicating a higher property jump at the shock front. The compressible vortex ring (CVR) is more pronounced exhibiting higher contrast and the CVR translation velocity is also higher. As the induced flow exits the shock tube, expansion fan structures are formed due to flow expansion, as shown in figure 9c. A trailing jet can be observed in case of higher charging voltages. In the case of the smaller channel (figure 9e,f), at a similar the Mach number, the translational velocity of the vortex is observed to be higher compared to the bigger channel.

At lower charging voltages, the CVR appears as a thin vortex loop indicating a very thin localized region having compressible effects near the vortex core similar to figure 9f. Multiple corrugations and oscillations were observed in the vortex ring as it travelled downstream. As the charging voltage is increased with the smaller channel, the velocity scales become significantly higher and the CVR appears even more pronounced with higher contrast. For high charging voltages, the CVR is observed to have embedded-shock structures. The figure also shows the presence of multiple smaller secondary vortices at the CVR (see figure 9e).

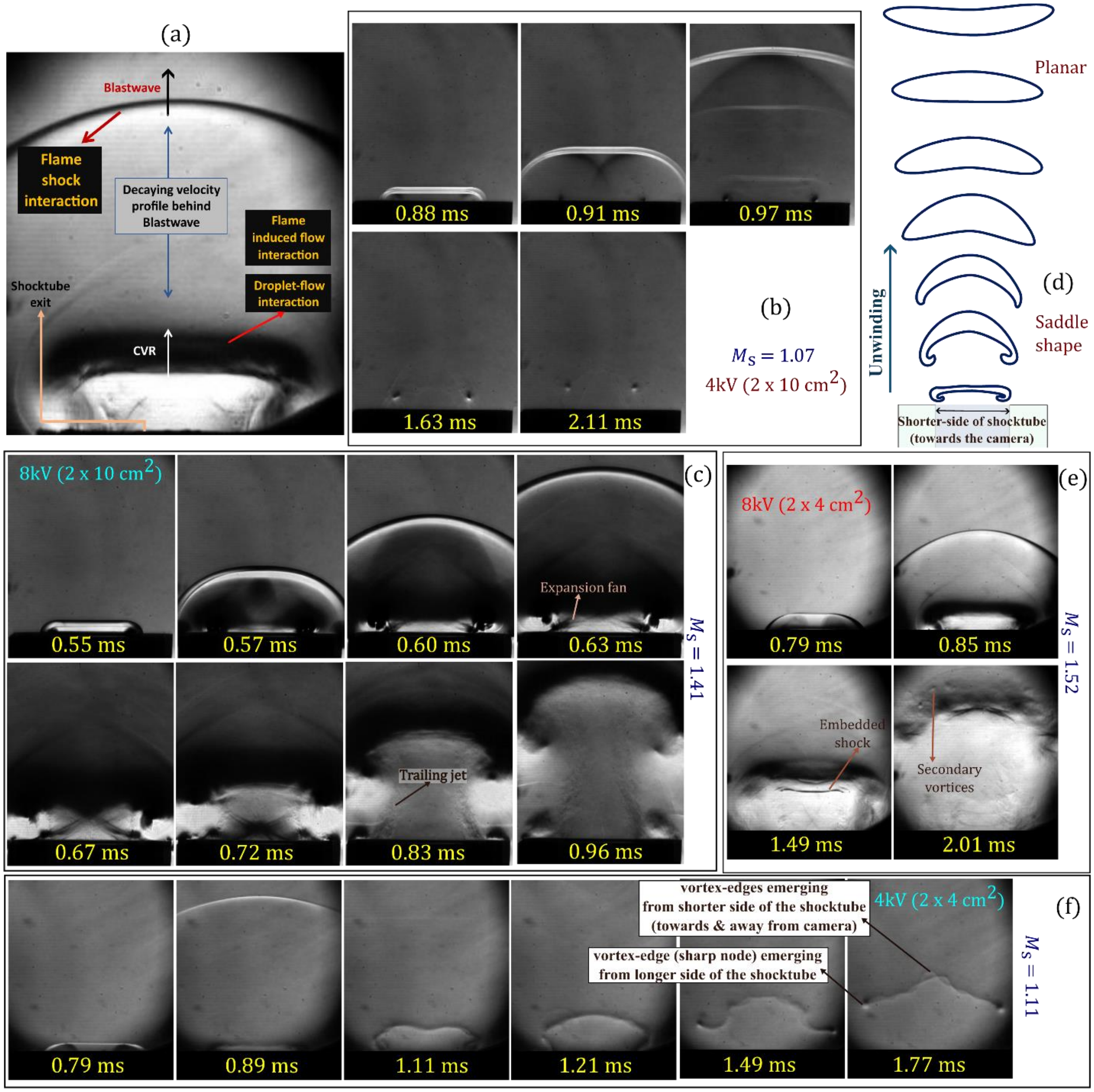

**Figure 9.** (a) Schematic showing the blastwave, induced flow behind it, and the compressible vortex ring (CVR) originating from the shock tube exit. Time series of the Schlieren imaging of the shock flow for different cases: (b) 4kV_B, (c) 8kV_B, (e) 4kV_S, (f) 8kV_S, (d) Schematic of the three-dimensional evolution of the compressible vortex ring (CVR) exiting the shock tube.

### 3.3.2. Simultaneous Parallax Schlieren and Mie-Scattering flow visualization:

It is to be noted that in all the experiments, the Schlieren imaging has been performed with the optical axis of the camera orthogonal to the plane of the shorter side of the rectangular shock tube channel. To get a better understanding of the vortical structures, simultaneous Schlieren and Mie-scattering flow visualization are performed. The Schlieren is set up with the optical axis orthogonal to the shorter side of the shock tube channel but at some parallax to obtain a better view of the three-dimensional structure of the flow features. A laser sheet is aligned parallel to the shorter side of the shock tube channel passing through the central plane and a high-speed camera is mounted with its optical axis orthogonal to the laser sheet. Thus, the horizontal x-direction in all the imaging is parallel to the shorter side of the shock tube and the longer side of the shock tube is along the perpendicular to the plane of the image. Figure 10 shows the time series of the CVR evolution obtained using simultaneous parallax Schlieren imaging and Mie-scattering flow visualization at the central plane parallel to the shorter side of the shock tube. The parallax Schlieren imaging clearly shows that the vortex loop has three-dimensional deformation. Since, both the imaging are obtained simultaneously, the macro features and the three-dimensional orientation of the CVR can be understood from the parallax Schlieren whereas, the local flow features at the central plane which interacts with the droplet can be obtained for the Mie-scattering imaging (length of the longer side is far greater than the droplet flame width).

The schematic of the evolution of the 3D macro features of the CVR is shown in figure 9d. For all the cases, the vortex ring is observed to attain a saddle-like shape with the leading edge of the saddle shape originating from the shorter side of the shock tube (towards and away from the camera) whereas the lagging edge of the saddle-shape originating from the longer side of the shock tube. This is evident in the parallax Schlieren images in figure 10a (right, at 2.00 ms), where the vortex loop towards and away from the camera (near the centre of the frame) are leading ahead whereas the sharp pointy features on the either side (left and right) lag behind (figure 9f). The central vortex loop edges arise from the shorter edges of the shock tube exit present towards and away from the camera, whereas the steep pointy edges on the either side arise from the longer edges of the shock tube exit which is parallel to the optical axis of the camera. It is evident from the figure 9d, figure 10b that the central horizontal edges of the vortex loop arising from the front and rear shorter edges of the shock tube exit (towards and away from the camera, see figure 10b) is slanted on the either side and these slanted front and rear edges meet at two points on either side (that appears as the sharp nodes in the parallax Schlieren imaging, shown in figure 10b). Thus, these sharp nodes at the junction of the slanted edges on the either side of the saddle-shaped vortex loop (emerging from both the shorter sides of shock tube) are located on the central plane of the shock tube (parallel to the shorter edge of the shock tube, i.e., plane of the laser sheet). This vortex behaviour matches with the experimental observations reported before in literature by Zare-Behtash et al. (2008, 2009) who studied the vortex loop evolution exiting a non-symmetrical nozzle. They showed that the vortex loop does not expand along the major axis as much as it expands along the minor axis and the portion of the vortex loop generated from the shorter edge has greater velocity downstream leaving behind the vortex loop portion from the longer side. This is reflected in the current experiments where the vortex edges emerging from longer side are lagging compared to that emerging from the shorter side.

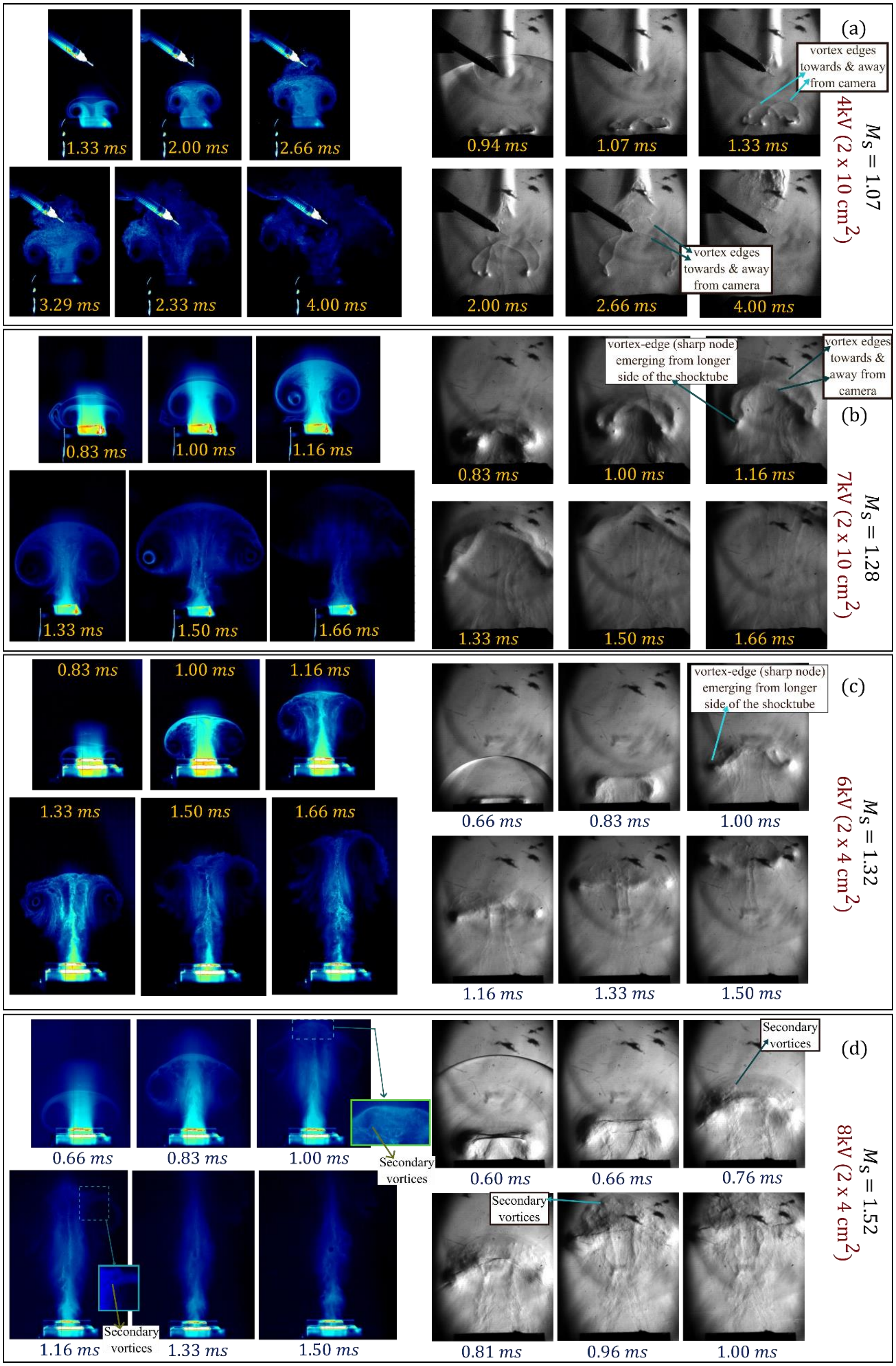

(a)
1.33 ms
2.00 ms
2.66 ms
3.29 ms
2.33 ms
4.00 ms
0.94 ms
1.07 ms
1.33 ms
2.00 ms
2.66 ms
4.00 ms
vortex edges towards & away from camera
vortex edges towards & away from camera
$M_S = 1.07$
4kV (2 x 10 cm$^2$)
(b)
0.83 ms
1.00 ms
1.16 ms
1.33 ms
1.50 ms
1.66 ms
0.83 ms
1.00 ms
1.16 ms
1.33 ms
1.50 ms
1.66 ms
vortex-edge (sharp node) emerging from longer side of the shocktube
vortex edges towards & away from camera
$M_S = 1.28$
7kV (2 x 10 cm$^2$)
(c)
0.83 ms
1.00 ms
1.16 ms
1.33 ms
1.50 ms
1.66 ms
0.66 ms
0.83 ms
1.00 ms
1.16 ms
1.33 ms
1.50 ms
vortex-edge (sharp node) emerging from longer side of the shocktube
$M_S = 1.32$
6kV (2 x 4 cm$^2$)
(d)
0.66 ms
0.83 ms
1.00 ms
Secondary vortices
1.16 ms
Secondary vortices
1.33 ms
1.50 ms
0.60 ms
0.66 ms
0.76 ms
Secondary vortices
Secondary vortices
0.81 ms
0.96 ms
1.00 ms
$M_S = 1.52$
8kV (2 x 4 cm$^2$)

**Figure 10.** Time series of the shock flow depicting the simultaneous planar Mie scattering flow visualization at the central plane parallel to the shorter edge of the shock tube (left) and the parallax Schlieren imaging (right): for bigger channel (2 cm × 10 cm c/s) at the charging voltages: (a) 4kV_B, (b) 7kV_B and for smaller channel (2 cm × 4 cm c/s) at the charging voltages; (c) 6kV_S, (d) 8kV_S.

As shown in figure 10, when the input energy is increased for the wire-explosion, the CVR translation velocity increases. The saddle-like shape of the CVR and the CVR evolution pattern as depicted in figure 9d is observed in all the cases in the parallax Schlieren images shown in figure 10. The central plane Mie-scattering visualization of the CVR shows a distinct, smooth boundary of the vortex structure without any instabilities till the case with $M_s$ < 1.30. However, beyond $M_s$ >1.30, the CVR boundary is observed to show KH instability-driven corrugations along the vortex boundary in the Mie-scattering imaging (figure 10e). At even higher energy input (8 kV), figure 10d (left) shows the presence of secondary vortices near the leading boundary of the CVR, that have formed due to the growth of KH instabilities.

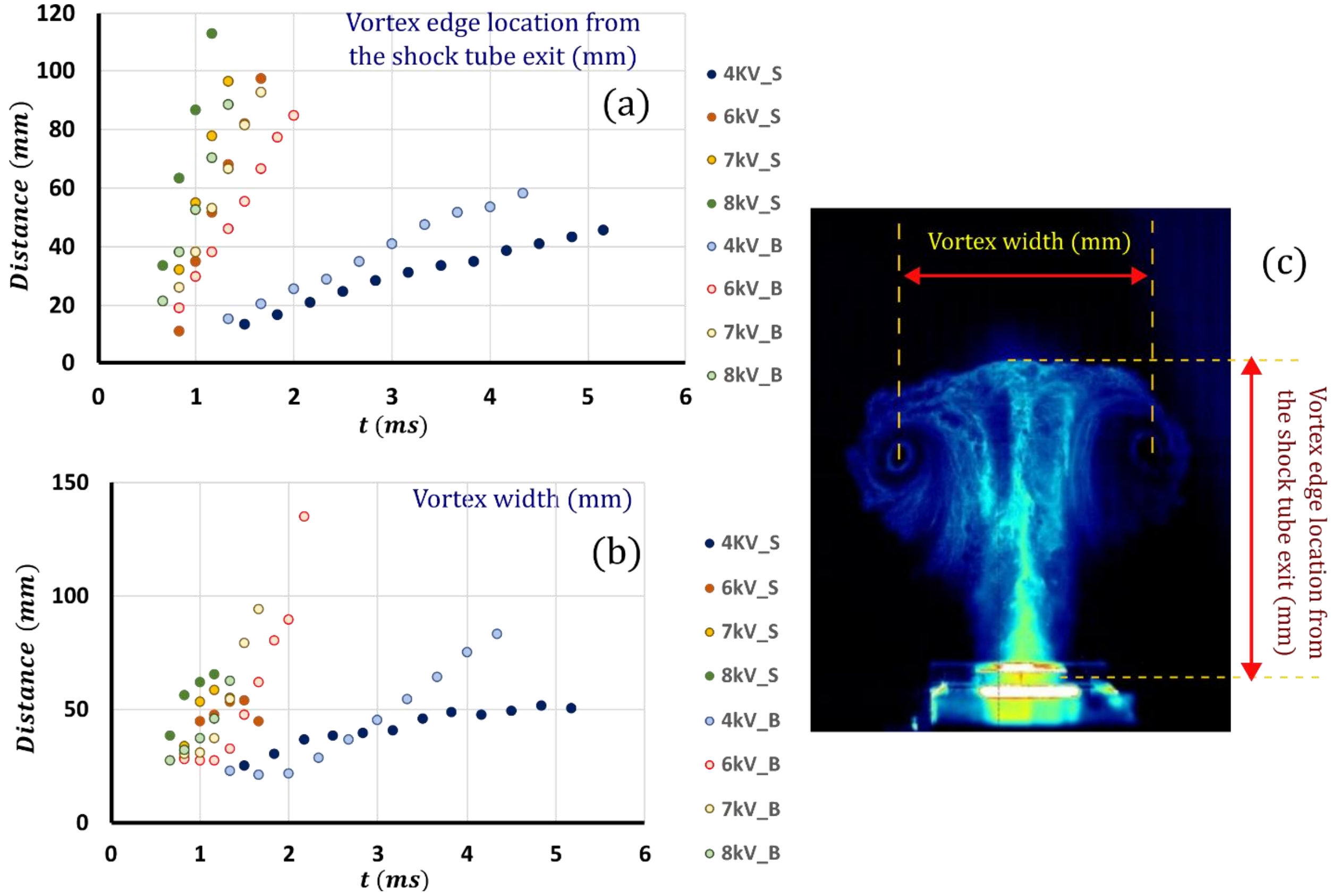

**Figure 11.** Plot of the temporal variation of the vortex (CVR) dimensions (mm): (a) vortex edge location and (b) vortex width (eye-to-eye distance) for all the cases of shock tube focusing. (c) Schematic showing the measurement of the dimensions of the CVR. Here, 'B' represents bigger channel (2 cm × 10 cm c/s), and 'S' represents smaller channel (2 cm × 4 cm c/s).

The pointy nodes in parallax Schlieren which are present on the central plane parallel to the shorter shock tube edge are tracked and are found to correspond with the vortex eyes that are observed in the planar Mie-scattering flow visualization in the central same plane. Figure 11 shows the plots of the characterization of the vortical structures arising from the shock tube for different cases. The CVR edge and vortex eye locations were spatially tracked with time and the vortex edge location, eye-to-eye distance are plotted with time in figure 11 a,b. It shows that the vortex translational velocity increases with an increase in input energy. It is observed that, the vortex width drastically increases after it exits the shock tube edge, and this expansion is more pronounced for lower charging voltages (4kV) as shown in figure 11b. The corresponding simultaneous parallax Schlieren imaging shows that the vortex loop is initially observed to be wound-up on the either side (see figure 9d), which gets unwound as the CVR propagates downstream away from the shock tube edge. It is observed that this unwinding phase corresponds to the rapid vortex expansion (increase in vortex width) that is observed in the planar vortex visualization in the central plane. This unwinding of the vortex loop can also be clearly observed in figure 9f from 1.11 – 1.77 ms for 4kV charging using a smaller shock tube. This unwinding is also seen for higher charging voltages (see 0.8 – 1.16 ms in figure 10f), however, the expansion rate is not as prominent as the 4 kV charging case. The time series of simultaneous parallax Schlieren and Mie scattering flow visualization with ignited droplet is shown in supplementary figure S3. It is to be noted that, since the amount of seeding and its distribution cannot be controlled (as explained in section 2), the Mie-scattering images are only used for evaluating the order of magnitude of the velocity scale associated with the vortex ($v_{CVR}$), which plays an important role in the droplet disintegration.

### 3.4. Flow – droplet interaction ($\tau > 1$)

The droplet dynamics during the interaction with the shock tube exhaust flow occur in two stages. During the initial stage of the interaction, continuous droplet deformation is observed which occurs in the time scale of the decay of the velocity imposed by blast wave ($v_s$), i.e., $\tau_s < \tau < 1$, as shown in figure 12a (green region). Due to the local velocity imposed by the blast wave ($v_s$), a high-pressure region is developed at the forward stagnation point of the droplet (polar location). This difference between stagnation pressure at the polar location and the static pressure at the equator of the droplet, i.e., ($\Delta P = P_p - P_{eq}$) causes deformation of the droplet into oblate shape, which results in continuous increase in the droplet length scale in the equatorial plane. The surface tension forces on the droplet surface resists this deformation. Hence, the non-dimensional parameter that indicates the competing forces involved in deformation i.e., the deformation Weber number can be written as, $We_d \sim \Delta P.R_e/\sigma$, which is the ratio of the deforming force and the restoring surface tension force, where $R_e$ is the equatorial radius of curvature (Sharma et al. 2023a). Hence, as the droplet starts to deform, the surface tension forces are overpowered by the deforming forces which is indicated by the increasing Weber number with deformation. Thus, the droplet continues to deform leading to flattening into oblate shape as shown in figure 12a,b (green region).

Furthermore, during the same time scale ($\tau_s < \tau < 1$), simultaneously alongside this continuous deformation, the KH instability-induced perturbations were also observed on the windward surface of the droplet at higher $M_s$ (higher instantaneous Weber numbers based on $v_s$), as shown in figure 5 and figure 12b (green region). These perturbations temporally grow into KH waves along the windward surface of the droplet leading to transport of the fluid towards the periphery in the equatorial plane. This results in formation of a sheet at the equatorial location. This similar mechanism has been reported by Sharma et al. 2021 which has been observed in current experiments, as shown in figure 12a (green region). It is to be noted that since the quartz rod is holding the droplet in pendant mode, due to the presence of the rod, the droplet breakup will not be same as that of the contactless droplet, however the qualitative response of the droplet dynamics will remain similar as long as the droplet dynamics does not enter the wake of the quartz rod. Since the initial deformation process, the KH wave formation and sheet formation are observed to occur upstream of the quartz rod, the droplet dynamics observed are qualitatively representative of that of contactless droplets.

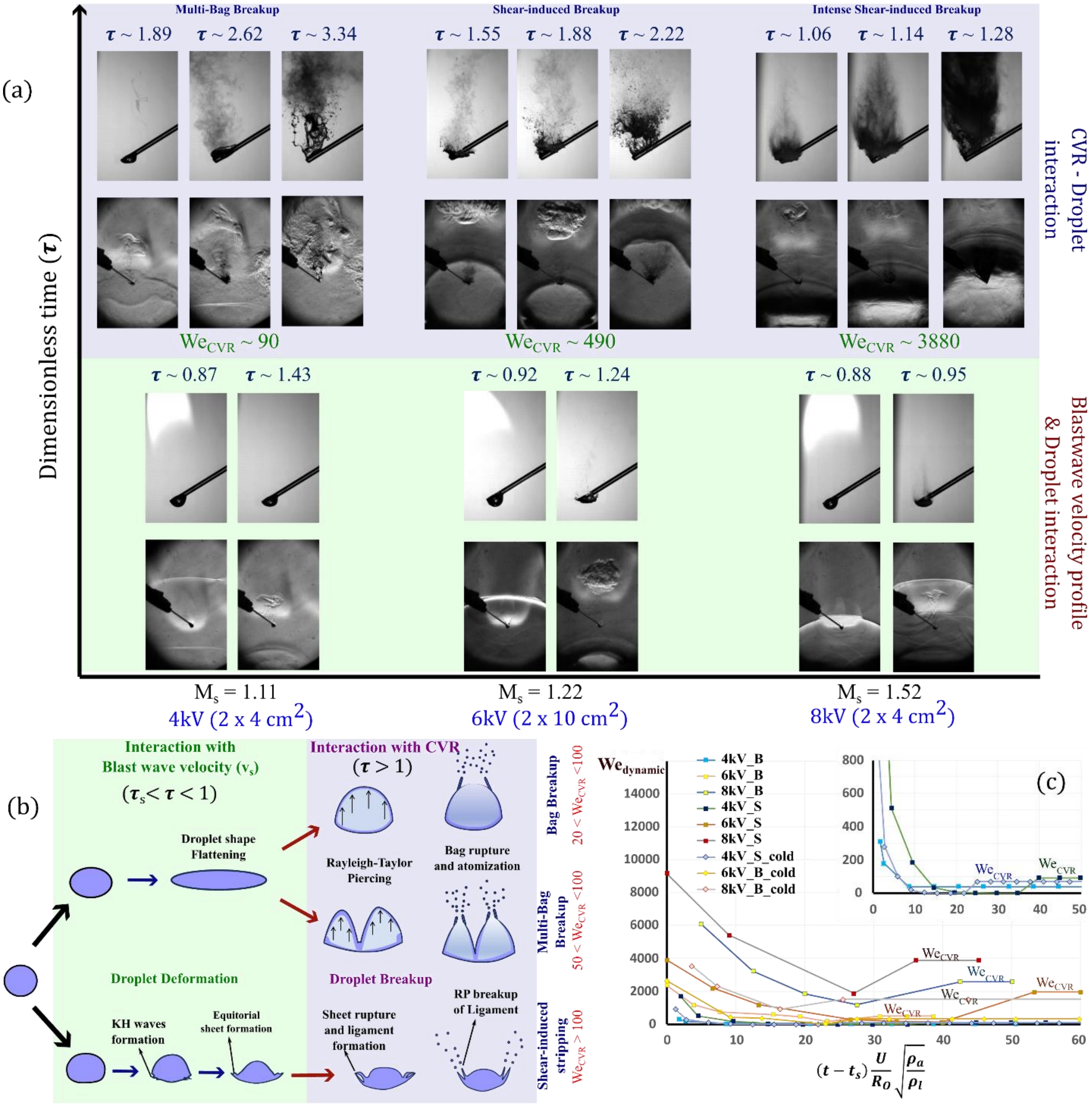

**Figure 12.** (a) Time series plot depicting two stages of the interaction with the droplet: 1. Decaying velocity profile behind the blast wave interacts with the droplet (green region) and 2. CVR formed from the induced flow exiting the shock tube interacting with the droplet (Purple region). The simultaneous time series images of the droplet shadowgraphy (top) and Schlieren (bottom) are shown in each of the interaction stages (with time stamps) for three different cases on the x-axis: 4kV_S, 6kV_B and 8kV_S respectively, (b) Schematic of the droplet breakup modes at different Weber numbers. (c) Dynamic temporal variation of Weber number ($We_{dynamic}$) due to temporally decaying velocity at the droplet behind the blast wave plotted against time (from shock interaction) normalized against inertial timescale of the droplet, plotted for different cases.

During this initial stage of droplet dynamics, due to $v_s$, the droplet starts to show deformation and KH instabilities (at high $M_s$). However, droplet breakup is not observed till the droplet starts to interact with the oncoming compressible vortex ring (CVR) formed from the induced flow behind the blast wave, which reaches the droplet after some delay (figure 12a, purple region). The Mie-scattering visualization of the induced flow CVR has been used to estimate the velocity scales associated with the CVR and it has been found that the velocity scale of the CVR ($v_{CVR}$) is smaller than the shock propagation velocity, however, it is sufficient to disintegrate the droplet. In the previous experiments investigating droplet – blast wave (wire explosion technique) interaction (Sharma et al. 2021), normal shock assumption has been considered for the planar blast wave (with decaying profile) for evaluating the Weber number. Even though the order of magnitude of the velocity scales matched with the Weber number range, the temporal variation in blastwave-imposed velocity and the subsequent induced flow has not been discussed in previous studies. Thus, the time scale of the two stages of the droplet interaction i.e., initial deformation ($\tau_s < \tau < 1$) and subsequent breakup ($\tau > 1$) has not been addressed. Thus, an attempt has been made here to incorporate the temporal variation of the velocity to evaluate the instantaneous Weber number at the droplet.The figure 12a shows the two stages of the flow – droplet interaction. Even though the droplet doesn't respond to the imposed blast wave, the velocity profile behind the blast wave ($v_s$) ensues droplet deformation and initiation of KH instabilities ($\tau_s < \tau < 1$). Subsequently, when the compressible vortex (CVR) reaches the droplet ($\tau > 1$), the instabilities intensify leading to the droplet fragmentation. The green region in figure 12a,b represents the initial interaction between the decaying velocity profile behind the blast wave ($v_s$) and the droplet. The purple region represents the interaction between the CVR and the droplet.

$$\mathrm{v}_{comb}(t) = \mathrm{v}_{s,Th}(t) + \mathrm{v}_{CVR}(t) \tag{3.21}$$

Where, $v_{comb}$ is the total combined velocity at the droplet due to the blast wave profile and CVR, which is plotted in figure 8a for different cases. $v_{s,Th}$ is the theoretical estimate of the instantaneous local velocity at the droplet imposed by the blast wave and $v_{CVR}$ is the velocity scale associated with the CVR. Before the vortex reaches the droplet, $v_{CVR} = 0$, hence $v_{comb} \sim v_{s,Th}(t)$. It is to be noted that by the time CVR reaches the droplet, the $v_s$ decays significantly and approaches zero and $v_{CVR} >> v_{s,Th}$, thus, $v_{comb} \sim v_{CVR}$ during the CVR – droplet interaction. Thus, the temporally varying Weber number ($We_{dynamic}$) is evaluated based on the instantaneous net local velocity imposed at the droplet, $v_{comb}(t)$, for different cases, which is given by:

$$\mathrm{We}_{dynamic}(t) = \frac{\rho d_i \mathrm{v}_{comb}(t)}{\sigma} \tag{3.22}$$

Where instantaneous droplet diameter ($d_i$) at the beginning of the interaction is taken as the characteristic length scale. ρ is the density of the impinging flow and σ is the surface tension of the liquid phase. The dynamic Weber number obtained from the above equation is plotted in figure 12c for different cases. It can be observed that $We_{dynamic}$ attains very high values initially which decay rapidly to lower values, and subsequently attains stepwise increase to a higher value ($We_{CVR}$) when the CVR reaches and starts to interact with the droplet (after some delay, $\tau > 1$). During the interaction with the decaying velocity of the blast wave because of the short time scales of the deforming forces on the droplet corresponding to the initial high $We_{dynamic}$ value, the droplet breakup is not observed to occur during this stage. However, this temporally varying dynamic Weber number ($We_{dynamic}$) is responsible for the droplet deformation and the initiation of KH instability during this initial stage. Subsequently, when the CVR arrives at the droplet ($\tau > 1$), the dynamic Weber number at the droplet ($We_{dynamic}$) experiences a jump to a higher value i.e., $We_{CVR}$, which is responsible for the final breakup of the droplet. $We_{CVR}$ has been labelled in figure 12c for different cases. Since the droplet breakup occurs only due to the interaction with the compressible vortex ring (CVR), the velocity scales ($V_i$) associated with the CVR can be used for characterizing the flow leading to the breakup. It is to be noted that due to the effect of the quartz rod. the later part of the droplet breakup is only qualitatively studied.

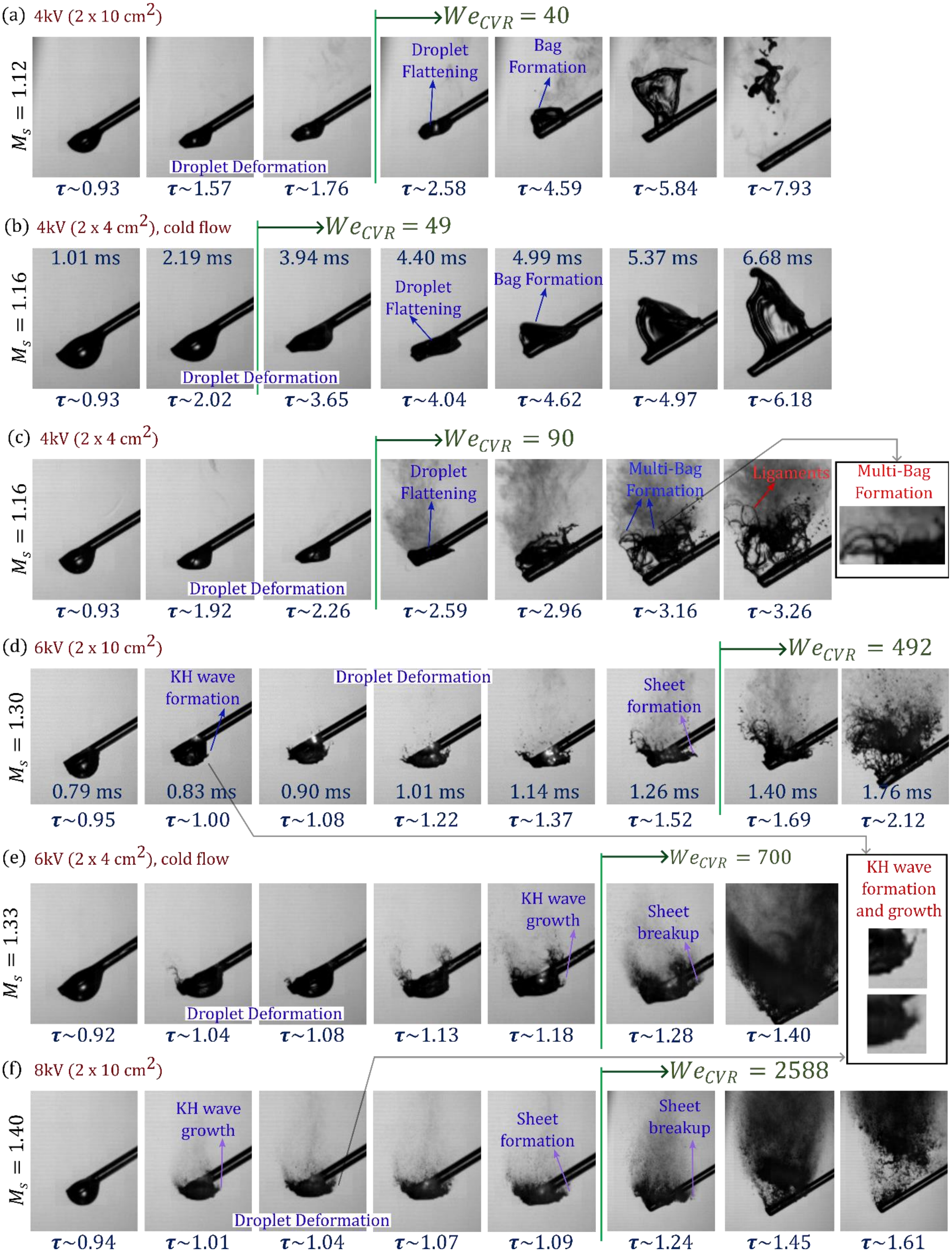

(a) 4kV (2 x 10 cm$^2$)
$We_{CVR} = 40$
$M_s = 1.12$
Droplet Flattening
Bag Formation
Droplet Deformation
$\tau$~0.93 $\tau$~1.57 $\tau$~1.76 $\tau$~2.58 $\tau$~4.59 $\tau$~5.84 $\tau$~7.93
(b) 4kV (2 x 4 cm$^2$), cold flow
$We_{CVR} = 49$
$M_s = 1.16$
1.01 ms 2.19 ms 3.94 ms 4.40 ms 4.99 ms 5.37 ms 6.68 ms
Droplet Flattening
Bag Formation
Droplet Deformation
$\tau$~0.93 $\tau$~2.02 $\tau$~3.65 $\tau$~4.04 $\tau$~4.62 $\tau$~4.97 $\tau$~6.18
(c) 4kV (2 x 4 cm$^2$)
$We_{CVR} = 90$
$M_s = 1.16$
Droplet Flattening
Multi-Bag Formation
Ligaments
Multi-Bag Formation
Droplet Deformation
$\tau$~0.93 $\tau$~1.92 $\tau$~2.26 $\tau$~2.59 $\tau$~2.96 $\tau$~3.16 $\tau$~3.26
(d) 6kV (2 x 10 cm$^2$)
$We_{CVR} = 492$
$M_s = 1.30$
KH wave formation
Droplet Deformation
Sheet formation
0.79 ms 0.83 ms 0.90 ms 1.01 ms 1.14 ms 1.26 ms 1.40 ms 1.76 ms
$\tau$~0.95 $\tau$~1.00 $\tau$~1.08 $\tau$~1.22 $\tau$~1.37 $\tau$~1.52 $\tau$~1.69 $\tau$~2.12
(e) 6kV (2 x 4 cm$^2$), cold flow
$We_{CVR} = 700$
$M_s = 1.33$
KH wave growth
Sheet breakup
KH wave formation and growth
Droplet Deformation
$\tau$~0.92 $\tau$~1.04 $\tau$~1.08 $\tau$~1.13 $\tau$~1.18 $\tau$~1.28 $\tau$~1.40
(f) 8kV (2 x 10 cm$^2$)
$We_{CVR} = 2588$
$M_s = 1.40$
KH wave growth
Sheet formation
Sheet breakup
Droplet Deformation
$\tau$~0.94 $\tau$~1.01 $\tau$~1.04 $\tau$~1.07 $\tau$~1.09 $\tau$~1.24 $\tau$~1.45 $\tau$~1.61

**Figure 13.** Shadowgraphy time series images of the droplet breakup dynamics during interaction with blast wave and induced flow for Combusting droplet: (a) 4kV_B, (c) 4kV_S, (d) 6kV_B, (f) 8kV_B and for Non-combusting droplet (cold flow): (b) 4kV_S, (e) 6kV_S. The CVR interaction occurs right side of the vertical green line and the corresponding $We_{CVR}$ is mentioned.

From figure 12a, it can be observed that the droplet breakup behaviour significantly changed as the Mach number is varied, where bag breakup mode is observed at lower Mach numbers, whereas, as $M_s$ is increased, the shear stripping of the droplet is observed. The following figure 13 shows the time series of the droplet breakup (with and without ignition) at different flow conditions obtained by varying the charging voltages in both the shock tube channels. The cases without droplet ignition are labelled as 'cold flow' in the figure. Figure 13a-c are the low Mach number cases where the velocity scales are slower corresponding to low Weber number CVR i.e., $We_{CVR}$ < 100, and figure 13d-f correspond to high $We_{CVR}$ cases. As discussed before, the droplet deformation is observed in all the cases. Figure 13 shows the two-stage droplet response to the imposed flow and the time instant from where CVR interaction (second stage, $\tau > 1$) starts is mentioned. It is to be noted that the increase in $M_s$ results in higher $v_{CVR}$ and this corresponds to a smaller time delay between blast wave incidence ($\tau \sim \tau_s$) and CVR interaction ($\tau > 1$). Thus, the induced flow reaches the droplet quicker closer to $\tau \sim 1^+$ as $M_s$ is increased which is also seen from figure 13 where CVR interacts with the droplet for $\tau > 1$ at smaller values of $\tau$ as $M_s$ is increased.

For low $We_{CVR}$ cases, the droplet deformation at $\tau_s < \tau < 1$ continues leading to the flattening of the droplet into a disc-like shape whose thickness continuously decreases due to liquid transport towards the periphery as the droplet is elongated in the equatorial direction. Subsequently, for $\tau > 1$, Rayleigh-Taylor piercing is observed which leads to blowup of the membrane into a bag-like structure, which eventually ruptures due to nucleation of holes in the thin membrane, leading to droplet disintegration and atomization. For $We_{CVR}$ > 50, the multi-bag Rayleigh-Taylor piercing is observed, and these droplet breakup behaviours are in good agreement with the literature (Sharma et al. 2023a). Figure 13a,c shows the droplet breakup for bigger and smaller channel for the same charging voltage of 4kV, which shows bag breakup in case of the bigger channel, multi-bag breakup in case of the smaller channel. This can be attributed to higher velocity scales achieved while focusing using smaller channel, which results in higher $We_{CVR}$.

In addition to this, at higher flow velocities (higher Weber number, i.e., $We_{CVR}$), along with the initial droplet deformation ($\tau_s < \tau < 1$), surface waves due to KH instability also starts to form on the windward surface of the droplet near the high shear forward stagnation region, as shown in figure 12b. As the KH waves travel radially outward and grow in amplitude, the fluid is carried outward which gets accumulated towards the periphery. When the KH waves reach the edge, they get deflected along the flow direction, undergoing flow entrainment leading to the formation of a thin sheet of the accumulated fluid at the periphery (see figure 13d-f). The thin sheet ruptures due to the nucleation of multiple holes which grow temporally leading to the breakup of the sheet into multiple ligaments. These ligaments undergo secondary atomization due to Rayleigh – Plateau instability forming numerous daughter droplets. This shear-induced breakup mode is observed to be dominant at higher Weber numbers as shown in figure 13d-f. At very high Weber numbers, the droplet breakup through shear stripping becomes more intense, forming a dense cloud of droplet spray. The shear induced stripping is dominant at high Weber numbers ($We_{CVR}$) due to faster growth rates of KH instability compared to droplet deformation, which leads to faster disintegration of the droplet through shear stripping at high $We_{CVR}$. The KH instability growth is observed to be occurring at shorter time scales (~ $O(10^{-1})$ ms) compared to the droplet deformation and Rayleigh-Taylor piercing mode time scale (~ $O(10^{0})$ ms), see figure 13b,d. Thus, as pointed out by Sharma et al. (2021), at lower $M_s$, i.e., lower Weber number, when velocity scales are low, the KH instability does not occur, and hence, the slower Rayleigh-Piercing mode of breakup due to continuous droplet deformation is observed. However, when $M_s$ is increased, due to higher velocity scales, KH waves start to form, leading to a shear stripping mode of breakup that occur at shorter time scales (~ O(1 $0^0$) ms), see figure 13b,d.

Interestingly, the droplet breakup phenomena are observed to change for the same flow conditions (shock tube and charging voltage) when the droplet is ignited (see figure 13b,c). This can be attributed to the higher droplet temperatures during combustion. In the absence of the flame, the droplet temperature is same as the ambient temperature, however, when the droplet is ignited, the droplet surface temperature will become wet bulb temperature due to the phase change occurring at the surface. Since sufficient time is given after the ignition before the shock interaction, it is reasonable to assume the droplet heat-up phase is completed and the bulk liquid in the droplet is at the wet bulb temperature, for simplicity. Thus, the fluid properties during the interaction with the vortex are at an elevated temperature which resulted in alteration of the droplet breakup phenomena at the same input flow conditions. Thus, the surface tension of the liquid droplet at its wet bulb temperature will be lower leading to higher Weber number values in the presence of the flame for the same flow conditions.

All these observations are in good agreement with the Weber number range ($We_{CVR}$) corresponding to classical breakup modes in the literature, which is valid for the droplet size is far greater than the order of magnitude of the wavelength of the KH wave, i..e, $d \gg \lambda$ (Sharma et al. 2021).

### 3.5. Regime Map

A regime map has been depicted in figure 14, showing the overall picture of the flame and droplet dynamics during the interaction between a combusting droplet and shock tube exhaust flow. The regime map depicts three stages of the interaction phenomena in current experiments, where the effect of shock Mach number is depicted for different cases for all the three stages of interaction. In figure 14, the cyan background is used to represent the Open-field blast wave case (without shock tube) whereas the red background is used to represent the shock tube focusing cases. Data points are plotted for different charging voltages and shock tube configurations with respect to the Shock Mach number. The representative images for those cases depicting the flame response are included in figure 14 accordingly. A time axis is shown in z-direction, depicting the time scale of occurrence of each stage of the interactions. The data points plotted in the regime map have different symbols with different colours to represent specific types of phenomena that occur at a specific stage of interaction. The regime map broadly is divided into two zones along the time axis: 1. interaction with the shock ($\tau_s < \tau < 1$) and 2. Interaction with the induced flow ($\tau > 1$). The direct shock wave interaction is only seen with the droplet flame whereas the droplet itself remains unaffected by it in the case of Open-field configuration. However, for the focused cases, the droplet shows deformation for all $M_s$ and simultaneous KH wave perturbation growth is also observed in case of high $M_s$. Furthermore, the induced flow in the open-field case affects only flame and the induced flow compressible vortex ring (CVR) that forms at the shock tube exit, only interacts with the droplet and not with the flame as the flame-blowout would've already occurred for focused cases by the time the CVR arrives. The induced flow – droplet interaction is plotted with respect to the Weber number based on the velocity scale of the induced flow ($v_{ind}$) instead of the $M_s$, in figure 14. For focused cases, the induced flow velocity scale is same as the velocity scale associated with CVR i.e., $v_{CVR}$.

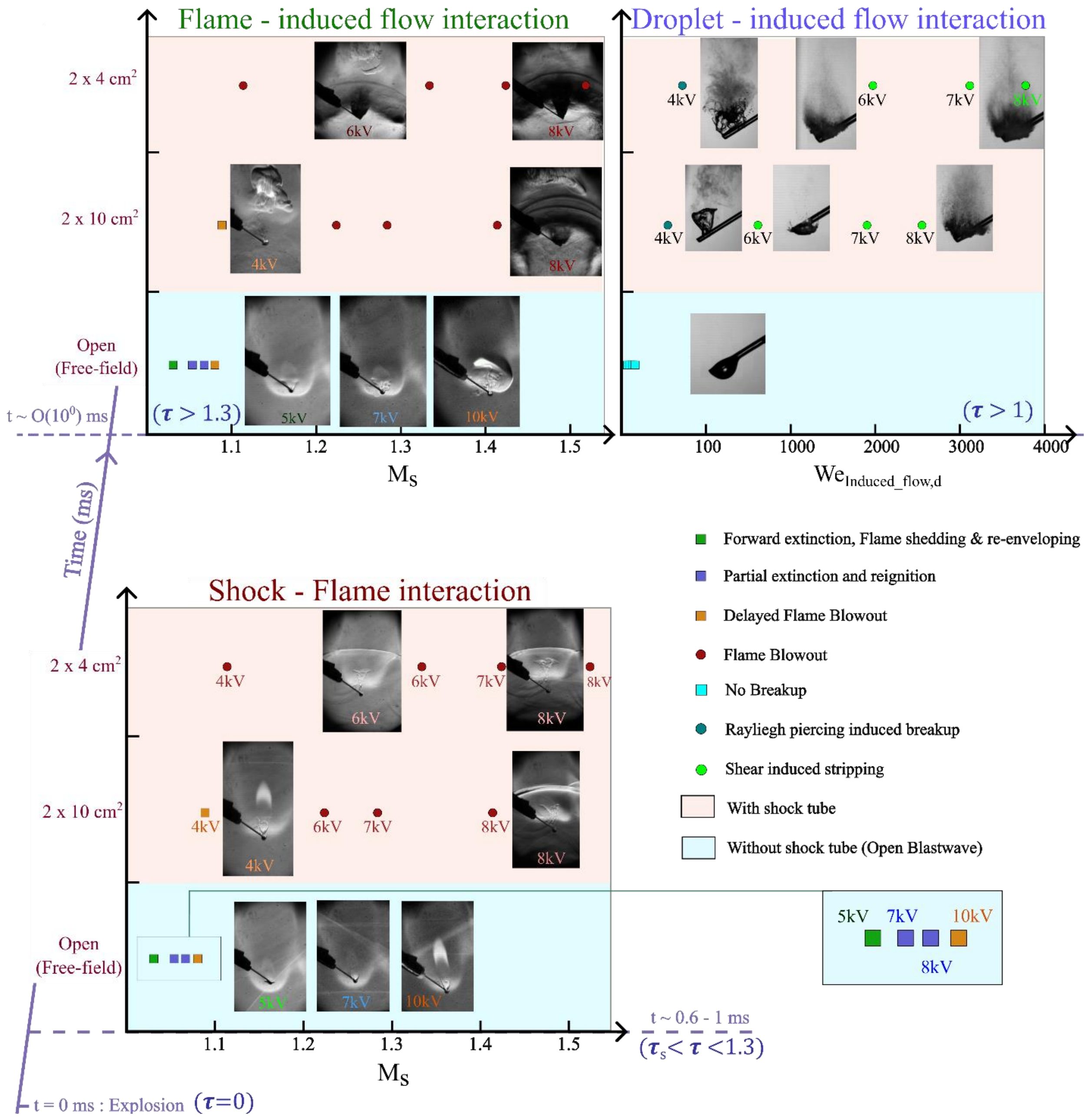

**Figure 14.** Regime map depicting phenomena in different configurations (Open-field and 2 shock tube focusing cases) in three interaction regimes: shock – flame, induced flow – flame, and induced flow CVR – droplet interactions with a time axis in the z-direction. The flame interaction regimes are plotted against $M_s$ whereas, the droplet interaction is plotted against the Weber number based on the induced flow and initial droplet diameter. The Cyan region represents the open-field and the Red region represents the shock tube focusing configurations respectively. All the different phenomena occurring in different regimes are represented by different symbols and color codes as shown in the legend.

## 4. Conclusion

The interaction between a combusting droplet and the flow imposed by a coaxially propagating blast wave is investigated experimentally with wire explosion-generated blast wave in both Open-field and shock tube-focused configurations. The addition of the shock tube directs and focuses the blast wave along the axial direction leading to higher Mach numbers ($M_s$), thus, facilitating a wide range of Mach numbers. An induced flow is observed behind the blast wave that reaches the droplet location after some time delay. The entire interaction with the droplet is found to occur in two stages: interaction with the velocity profile ($v_s$) behind the blast wave ($\tau_s < \tau < 1$) and the interaction with the induced flow ($v_{ind}$) behind the blast wave ($\tau > 1$). The droplet flame is observed to interact with the blast wave velocity profile and subsequently with the induced flow showing two stages of response in two different time scales. The theoretical model for blast wave propagation obtained with power-law density profile assumption is used to obtain the velocity profiles imposed by the blast wave in open-field configuration. Furthermore, the blast wave emerging from the shock tube opening is modelled to transition from planar to a cylindrical blast wave as it propagates downstream, which agrees with the experiments. However, this theoretical variation of local velocity at the droplet location is only valid during the initial stages, after which the entrainment effects start to affect the flow, which leads to induced flow. The local velocity at the droplet is initially due to the temporally decaying velocity profile behind the blast wave ($v_s$), which eventually approaches near zero and subsequently, the induced flow ($v_{ind}$) due to the entrainment effects reaches the droplet. The relatively slower induced flow is observed to form a compressible vortex ring (CVR) in a shock tube focused configuration, as it exits the shock tube. The vortex dynamics and shape evolution has been studied and the velocity scales involved were experimentally measured, to get a comprehensive understanding of the flow qualitatively and quantitatively.

The droplet and flame are observed to respond to the imposed flow independently. The flame is observed to interact with the blast wave velocity profile ($v_s$) during the initial stages which results in forward extinction, flame lift-off, which leads to extinction for higher $M_s$. It has been shown that the flame lift-off directly responds to the velocity profile behind the blast wave. For $M_s > 1.1$, the flame is observed to fully extinguish during the initial interaction with the blast wave. However, for lower $M_s$, the flame survives beyond the initial blast wave interaction, and it starts to interact with the slower induced flow ($v_{ind}$) subsequently, at longer time scales. The flame is observed to interact with the induced flow ($v_{ind}$) exhibiting a wide range of responses such as forward extinction, shedding, lift-off, partial extinction, and reignition depending on the Mach number. The different sub-regimes of the flame behavior have been identified and flame extinction criteria, flame shedding criteria have been proposed based on the flow imposed. The flame base shape evolution during the initial interaction with the blast wave has been explained using RM instability. The droplet interaction with the blast wave and the induced flow has also been investigated which showed only showed effect on the droplet in focused configurations. The droplet exhibited temporal deformation into an oblate shape for all these cases and the droplet dynamics were observed to occur in two stages: Initial interaction with the blast wave and subsequent interaction with the CVR. At lower $M_s$, the droplet elongates equatorially and undergoes Rayleigh-Taylor piercing exhibiting bag breakup mode of atomization during the eventual interaction with induced flow vortex (CVR). Unlike lower $M_s$, alongside continuous deformation, the droplet exhibited KH instability-induced perturbation growth on the windward surface. This leads to shear induced stripping as the perturbations further grow leading to sheet formation and rupture at the equator resulting in secondary atomization. The temporally varying dynamic Weber number has been estimated based on the temporally varying flow which gives further insights into the droplet dynamics. The Weber number ranges based on the induced flow velocity scale ($v_{ind}$) for different modes were observed to be in good agreement with the literature.

**Acknowledgements**

The authors are thankful to SERB (Science and Engineering Research Board) - CRG: CRG/2020/000055 for financial support. S.B. acknowledges funding through the Pratt and Whitney Chair Professorship.

# Insights into spatio-temporal dynamics during shock – droplet flame interaction

Gautham Vadlamudi [1], Akhil Aravind [1], Saini Jatin Rao [1], and Saptarshi Basu [1,2]*

*: Corresponding author email: sbasu@iisc.ac.in

**Affliations:**

[1] Department of Mechanical Engineering, Indian Institute of Science, Bangalore 560012, India

[2] Interdisciplinary Centre for Energy Research (ICER), Indian Institute of Science, Bangalore 560012, India

### Shock Arrival time theoretical vs experimental

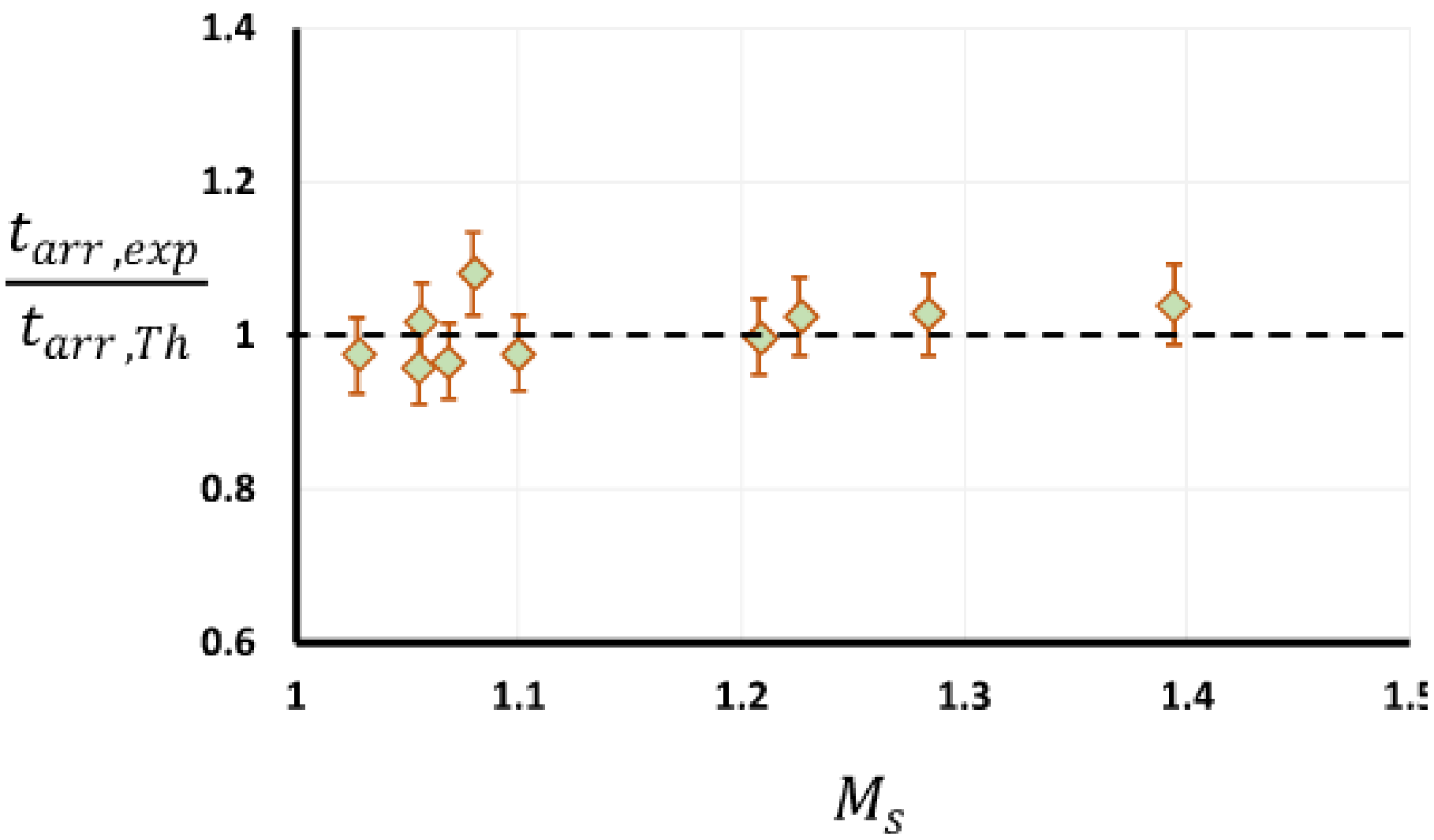

**Figure S1.** The ratio of experimentally obtained shock arrival time at the droplet location ($t_{arr,exp}$) to the theoretically estimated shock arrival time at the droplet location ($t_{arr,Th}$) is plotted against the corresponding shock Mach number.

The shock arrival time at the droplet location obtained from experiments ($t_{arr,exp}$) i

The shock arrival time at the droplet location is obtained theoretically ($t_{arr,Th}$) using Equation 3.11 in main manuscript, by substituting the characteristic length scale ($R_o$) that has been obtained by iteratively matching the theoretical estimate of shock Mach number ($M_s$) at the droplet location (Bach and Lee 1970) to the experimental value. The plot in Figure S1 shows that the shock arrival time of current experiments is in good agreement with the theory.

**Temporal pressure variation showing the time scale ($t_d$) which is used to obtain normalized time**

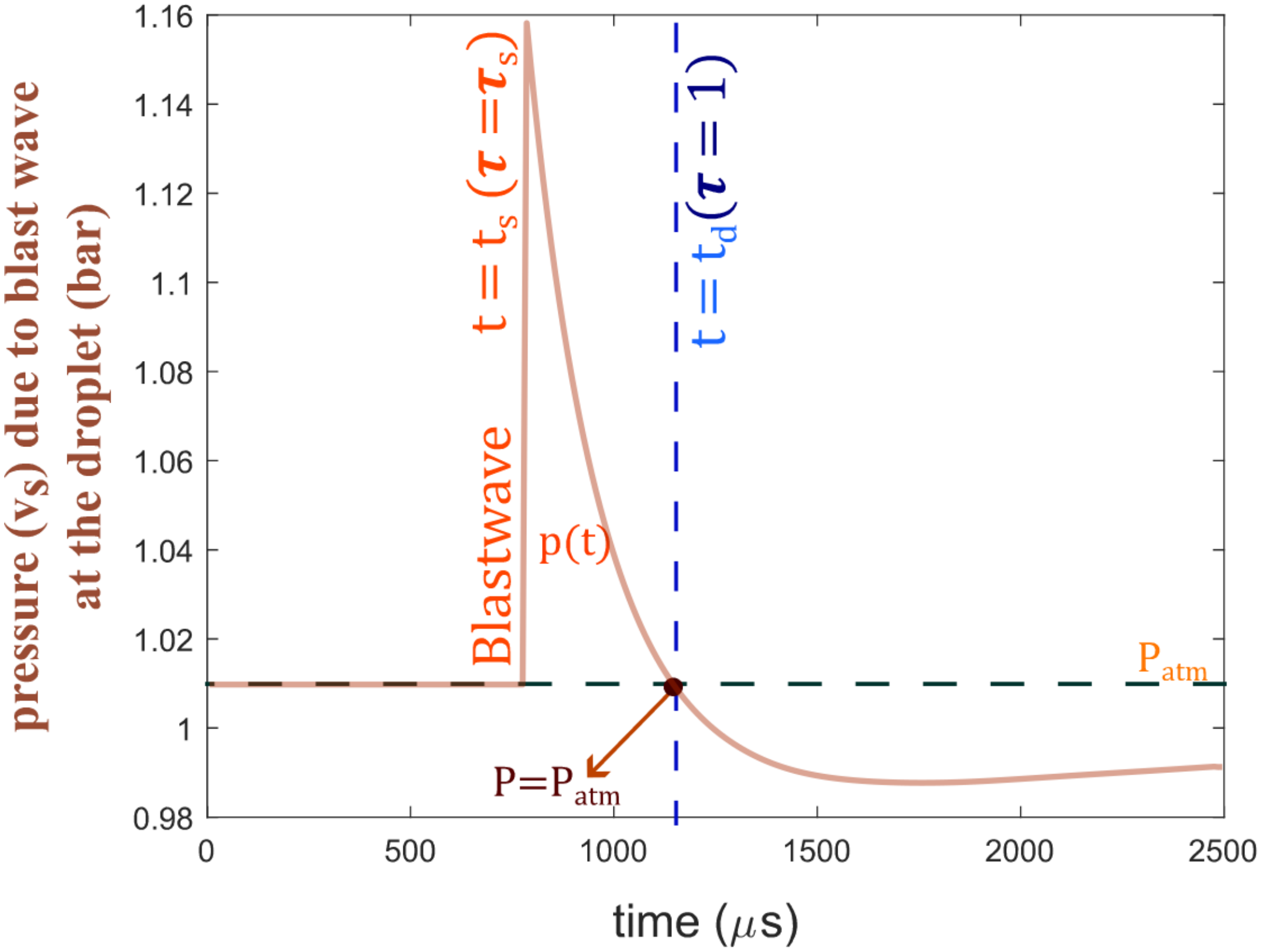

**Figure S2.** The theoretically obtained pressure variation at the droplet location with respect to time are plotted. t = 0 corresponds to the instant of explosion and $t = t_s$ represents the time instant the blast wave reaches the droplet location for 10kV_Open case ($M_s \sim 1.065$). The time instant where the decaying pressure at the droplet location reaches atmospheric pressure ($P_{atm}$) is denoted by $t = t_d$.

The induced flow ($v_{ind}$) due to the entrainment effects occurs due to the low pressure ($P < P_{atm}$) developed locally, behind the blast wave, the time scale at which the local pressure decays below the ambient ($P_{atm}$) i.e., $t_d$ is relevant to determine the two stages of interaction: interaction with the blast wave profile velocity ($v_s$) and interaction with the induced flow ($v_{ind}$). Thus, the time from explosion is normalized using the time scale '$t_d$' and the dimensionless time is given by:

$$\tau = \frac{t}{t_d}$$

This dimensionless time ($\tau$) becomes unity when the local pressure at the droplet decays and reaches $P_{atm}$ and induced flow ($v_{ind}$) is only observed for $\tau > 1$ and thus, for $\tau < 1$, only velocity scale that affects is imposed on the droplet-flame is the decaying velocity profile of the blast wave ($v_s$). The induced flow velocity scale ($v_{ind}$) is observed to be significantly smaller compared to the shock velocity scales, which results in a time delay for the induced flow to reach the droplet location after the shock interaction ($t = t_s$). The induced flow velocity scale ($v_{ind}$) is observed to increase with increase in shock strength ($M_s$) and depending on the $M_s$, the induced flow reaches the droplet at different times and for open-field configuration, the induced flow effects are observed after $\tau > 1.3$ and for focused cases ($M_s > 1.1$) the induced flow CVR reaches the droplet after $\tau > 1$.

**Simultaneous Mie scattering flow visualization and Parallax Schlieren for hot flow (with flame)**

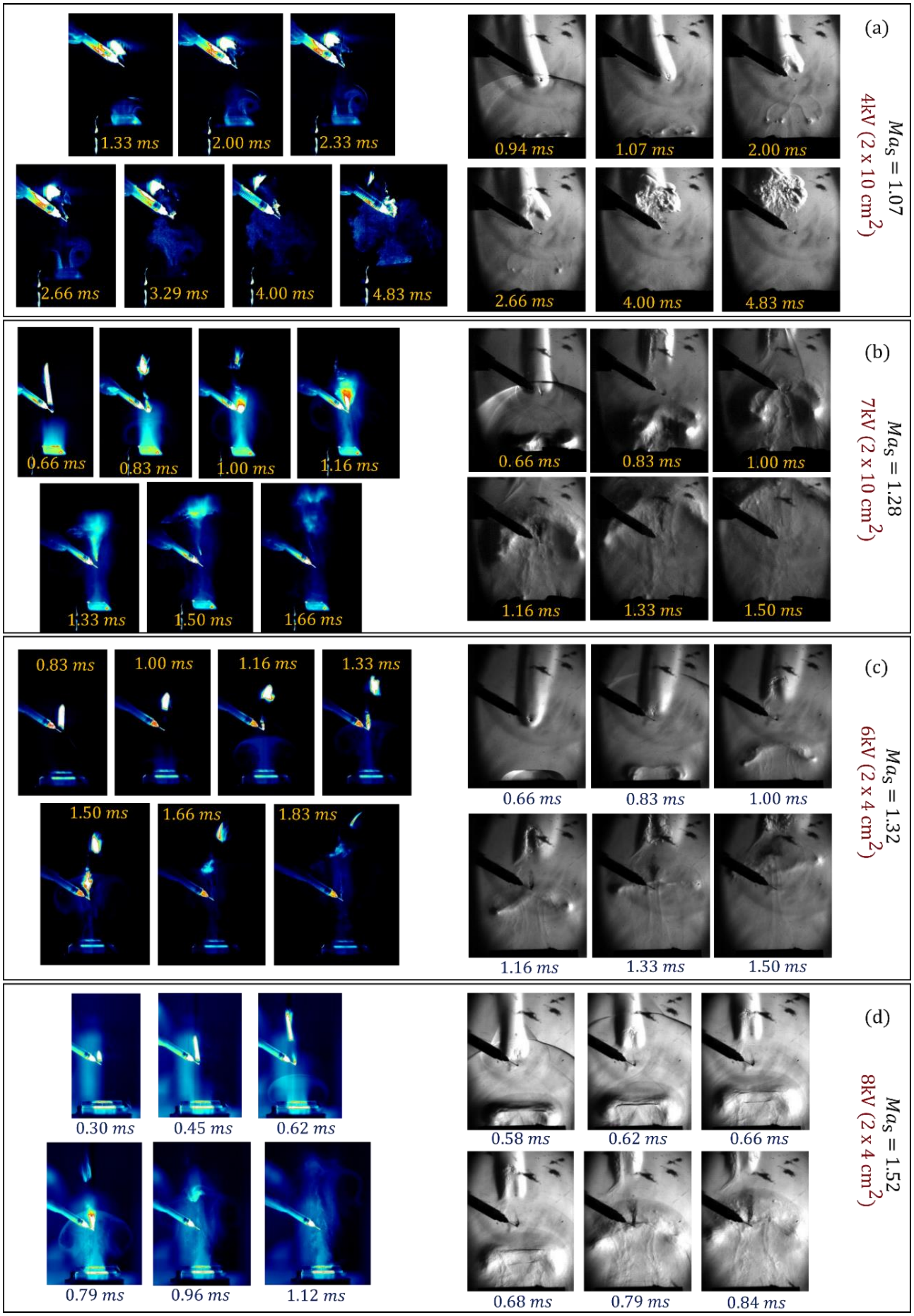

**Figure S3.** Time series of the interaction of the shock flow with the droplet flame, depicting the simultaneous planar Mie scattering flow visualization at the central plane parallel to the shorter edge of the shock tube (left) and the parallax Schlieren imaging (right): for larger channel (2 cm × 10 cm c/s) at the charging voltages: (a) 4kV, (b) 7kV and for smaller channel (2 cm × 4 cm c/s) at the charging voltages; (c) 6kV, (d) 8kV.

Figure 11 in main manuscript just shows the temporal evolution of the compressible vortex ring (CVR) exiting the shock tube. Figure S2 shows simultaneous parallax Schlieren and Mie-scattering imaging of the interaction of the CVR with the droplet flame. It can be clearly seen that the flame blowout already occurs in all the cases till the compressible vortex (CVR) starts to interact with the droplet flame. It is to be noted that for the Mie-scattering flow visualization, the seeding is introduced into the shock tube prior to the experiment and the seeding particles are observed to only exit the shock tube along with the CVR and not with the shock wave or the induced flow behind the shock wave. Furthermore, the Mie-scattering flow visualization is recorded using a high-speed camera by using a 10 μs exposure width of the laser sheet illumination sufficient to freeze the high-speed phenomena of the experiments. However, the camera recording is done at a lower acquisition rate (6000 fps) that corresponds to a larger exposure time in comparison with the shock flow time scales in current experiments. Thus, the features that are visible through the laser illumination scattering are frozen and captured at a very low exposure time, however, illuminating sources other than the synced laser such as the droplet flame will be captured at a longer exposure time corresponding to the acquisition rate, which results in motion blur for the flame visuals. This can be observed in the first few images in each case (before the vortex arrives) in figure S2, where the flame appears to be stretched to drastic lengths, however, that apparent elongation is a visual artifact of the motion blurring of the flame blowout phenomenon that occurs during the first stage of the interaction between shock wave ($\tau < 1$) and the droplet flame. This is also evident from the high-speed Schlieren and high-speed OH* Chemiluminescence imaging conducted at higher fps (smaller exposure times) that avoids motion blur. For $M_s$ > 1.1, the flame is observed to blow out and fully extinguish before the CVR arrives at the droplet. Furthermore, the droplet breakup is primarily occurring during the interaction with the CVR which is evident from figure S2. It is also observed that after the droplet breakup the droplet spray cloud (daughter droplets) gets entrained in the CVR (compressible vortex ring). In addition to the Mie-scattering imaging which are obtained using single-pulse mode of the high-speed laser, the data has also been captured using the double-pulse laser with inter pulse spacing of 10 μs, which allows to record consecutive frames of the compressible vortex propagation with very small time-delay. Since uniform seeding of the shock tube flow is experimentally challenging due to its intrinsic transient nature, performing PIV algorithm on these consecutive frames could not give a detailed spatial flow-field. However, the order of magnitude of the velocity scale associated with the CVR along the centerline (where the droplet is located) can be obtained from the vectors generated using the PIV algorithm. These values are found to be of similar order as that of the CVR translational velocity that is obtained from the Schlieren imaging.

**Figure S4: Logic gate circuit to alter the clock-pulse signal from camera to have intermittent dim frame to simultaneously visualize the flame location and tip.**

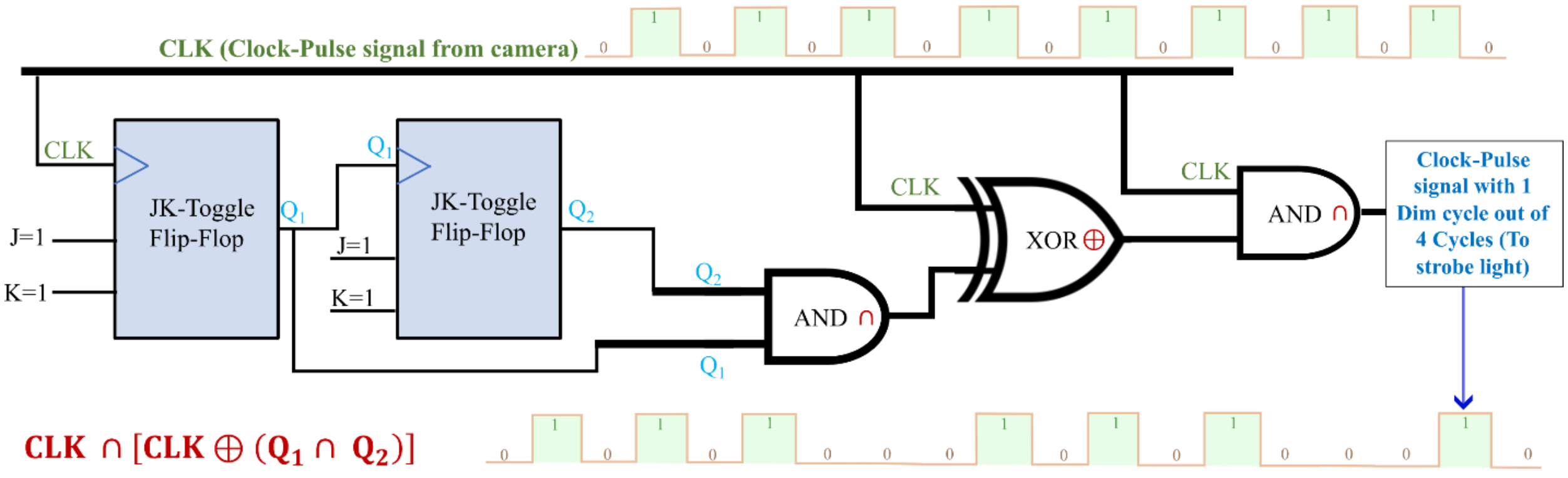

Figure S4 shows the Logic circuit used to convert the clock pulse signal from the camera (syncing) to intermittent dimming signal (1 out of 4 cycles is dim), to image the flame by making it better visible in the dark frame (no back light) to simultaneously get the flame location information during Schlieren imaging (for reference). This flame information is corroborated by the simultaneous high-speed OH* Chemiluminescence imaging to get better understanding about the flow around the flame along with the flame location and dimensions. The Logic circuit consists of two JK-type Toggle flip-flops (in series) and a combination of logic gates to obtain the desired signal.